\documentclass[twocolumn,prc,showpacs,showkeys,nofootinbib,preprintnumbers,superscriptaddress,amsmath,amssymb,nofootinbib]{revtex4}

\usepackage{graphicx}
\usepackage{dcolumn}
\usepackage{bm}
\begin{document}


\title{The $^{150}$Nd($^3$He,$t$) and $^{150}$Sm($t$,$^3$He) reactions with applications to $\beta\beta$ decay of $^{150}$Nd.}

\author{C.J. Guess} \email{carol.guess@gmail.com}
\affiliation{National Superconducting Cyclotron Laboratory, Michigan State University, East Lansing, MI 48824-1321, USA}
\affiliation{Department of Physics and Astronomy, Michigan State University, East Lansing, MI 48824, USA}
\affiliation{Joint Institute for Nuclear Astrophysics, Michigan State University, East Lansing, MI 48824, USA}

\author{T. Adachi}
\altaffiliation[Present Address: ]{KVI, University of Groningen, NL-9747 AA Groningen, The Netherlands}
\affiliation{Research Center for Nuclear Physics, Osaka University, Ibaraki, Osaka 567-0047, Japan}

\author{H. Akimune}
\affiliation{Department of Physics, Konan University, Okamoto 8-9-1, Higashinada, Kobe 658-8501, Japan}

\author{A. Algora}
\affiliation{Instituto de F\'{i}sica Corpuscular, CSIC-Universidad de Valencia, E-46071 Valencia, Spain}
\affiliation{Institute of Nuclear Research of the Hungarian Academy of Sciences, Debrecen H-4001, Hungary}

\author{Sam M. Austin}
\affiliation{National Superconducting Cyclotron Laboratory, Michigan State University, East Lansing, MI 48824-1321, USA}
\affiliation{Joint Institute for Nuclear Astrophysics, Michigan State University, East Lansing, MI 48824, USA}

\author{D. Bazin}
\affiliation{National Superconducting Cyclotron Laboratory, Michigan State University, East Lansing, MI 48824-1321, USA}

\author{B.A. Brown}
\affiliation{National Superconducting Cyclotron Laboratory, Michigan State University, East Lansing, MI 48824-1321, USA}
\affiliation{Department of Physics and Astronomy, Michigan State University, East Lansing, MI 48824, USA}
\affiliation{Joint Institute for Nuclear Astrophysics, Michigan State University, East Lansing, MI 48824, USA}

\author{C. Caesar}
\altaffiliation[Present address: ]{GSI Darmstadt, Helmholtz-Zentrum f\"{u}r Schwerionenforschung, D-64291, Darmstadt, Germany}
\affiliation{National Superconducting Cyclotron Laboratory, Michigan State University, East Lansing, MI 48824-1321, USA}
\affiliation{Johannes Gutenberg Universit\"{a}t, D-55099 Mainz, Germany}

\author{J.M. Deaven}
\affiliation{National Superconducting Cyclotron Laboratory, Michigan State University, East Lansing, MI 48824-1321, USA} \affiliation{Department of Physics and Astronomy, Michigan State University, East Lansing, MI 48824, USA}
\affiliation{Joint Institute for Nuclear Astrophysics, Michigan State University, East Lansing, MI 48824, USA}

\author{H. Ejiri}
\affiliation{Research Center for Nuclear Physics, Osaka University, Ibaraki, Osaka 567-0047, Japan}
\affiliation{Nuclear Science, Czech Technical University, Brehova, Prague, Czech Republic}

\author{E. Estevez}
\affiliation{Instituto de F\'{i}sica Corpuscular, CSIC-Universidad de Valencia, E-46071 Valencia, Spain}

\author{D. Fang}
\affiliation{Institut f\"ur Theoretische Physik, Universit\"at T\"ubingen, Auf der Morgenstelle 14, D-72076 T\"ubingen, Germany}

\author{A. Faessler}
\affiliation{Institut f\"ur Theoretische Physik, Universit\"at T\"ubingen,  Auf der Morgenstelle 14, D-72076 T\"ubingen, Germany}

\author{D. Frekers}
\affiliation{Institut f\"{u}r Kernphysik, Westf\"{a}lische Wilhelms-Universit\"{a}t M\"{u}nster, D-48149 M\"{u}nster, Germany}

\author{H. Fujita}
\affiliation{Research Center for Nuclear Physics, Osaka University, Ibaraki, Osaka 567-0047, Japan}

\author{Y. Fujita}
\affiliation{Department of Physics, Osaka University, Osaka 560-0043, Japan}

\author{M. Fujiwara}
\affiliation{Research Center for Nuclear Physics, Osaka University, Ibaraki, Osaka 567-0047, Japan}
\affiliation{Quantum Beam Science Directorate, Japan Atomic Energy Agency, Tokai, Ibaraki, 319-1195, Japan}

\author{G.F. Grinyer}
\altaffiliation[Present address: ]{Grand Acc\'{e}l\'{e}rateur National d'Ions Lourds (GANIL), CEA/DSM-CNRS/IN2P3, Bvd Henri Becquerel, 14076 Caen, France}
\affiliation{National Superconducting Cyclotron Laboratory, Michigan State University, East Lansing, MI 48824-1321, USA}

\author{M.N. Harakeh}
\affiliation{Kernfysisch Versneller Instituut, University of Groningen, NL-9747 AA Groningen, The Netherlands}

\author{K. Hatanaka}
\affiliation{Research Center for Nuclear Physics, Osaka University, Ibaraki, Osaka 567-0047, Japan}

\author{C. Herlitzius}
\altaffiliation[Present address: ]{Physics Department, Technische Universit\"{a}t M\"{u}nchen, D-85748 Garching, Germany}
\affiliation{National Superconducting Cyclotron Laboratory, Michigan State University, East Lansing, MI 48824-1321, USA}
\affiliation{Johannes Gutenberg Universit\"{a}t, D-55099 Mainz, Germany}

\author{K. Hirota}
\affiliation{Research Center for Nuclear Physics, Osaka University, Ibaraki, Osaka 567-0047, Japan}

\author{G.W. Hitt}
\altaffiliation[Present address: ]{Khalifa University of Science, Technology \& Research, P.O. Box 127788, Abu Dhabi, UAE}
\affiliation{National Superconducting Cyclotron Laboratory, Michigan State University, East Lansing, MI 48824-1321, USA} \affiliation{Department of Physics and Astronomy, Michigan State University, East Lansing, MI 48824, USA}
\affiliation{Joint Institute for Nuclear Astrophysics, Michigan State University, East Lansing, MI 48824, USA}

\author{D. Ishikawa}
\affiliation{Research Center for Nuclear Physics, Osaka University, Ibaraki, Osaka 567-0047, Japan}

\author{H. Matsubara}
\affiliation{Research Center for Nuclear Physics, Osaka University, Ibaraki, Osaka 567-0047, Japan}

\author{R. Meharchand}
\affiliation{National Superconducting Cyclotron Laboratory, Michigan State University, East Lansing, MI 48824-1321, USA}
\affiliation{Department of Physics and Astronomy, Michigan State University, East Lansing, MI 48824, USA}
\affiliation{Joint Institute for Nuclear Astrophysics, Michigan State University, East Lansing, MI 48824, USA}

\author{F. Molina}
\affiliation{Instituto de F\'{i}sica Corpuscular, CSIC-Universidad de Valencia, E-46071 Valencia, Spain}

\author{H. Okamura}
\altaffiliation{Deceased}
\affiliation{Research Center for Nuclear Physics, Osaka University, Ibaraki, Osaka 567-0047, Japan}

\author{H.J. Ong}
\affiliation{Research Center for Nuclear Physics, Osaka University, Ibaraki, Osaka 567-0047, Japan}

\author{G. Perdikakis}
\affiliation{National Superconducting Cyclotron Laboratory, Michigan State University, East Lansing, MI 48824-1321, USA} \affiliation{Joint Institute for Nuclear Astrophysics, Michigan State University, East Lansing, MI 48824, USA}

\author{V. Rodin}
\affiliation{Institut f\"ur Theoretische Physik, Universit\"at T\"ubingen, Auf der Morgenstelle 14, D-72076 T\"ubingen, Germany}

\author{B. Rubio}
\affiliation{Instituto de F\'{i}sica Corpuscular, CSIC-Universidad de Valencia, E-46071 Valencia, Spain}

\author{Y. Shimbara}
\affiliation{Graduate School of Science and Technology, Niigata University, Niigata 950-2181, Japan}

\author{G. S\"{u}soy}
\affiliation{Istanbul University, Faculty of Science, Physics Department, 34134, Vezneciler/Istanbul}

\author{T. Suzuki}
\affiliation{Research Center for Nuclear Physics, Osaka University, Ibaraki, Osaka 567-0047, Japan}

\author{A. Tamii}
\affiliation{Research Center for Nuclear Physics, Osaka University, Ibaraki, Osaka 567-0047, Japan}

\author{J.H. Thies}
\affiliation{Institut f\"{u}r Kernphysik, Westf\"{a}lische Wilhelms-Universit\"{a}t M\"{u}nster, D-48149 M\"{u}nster, Germany}

\author{C. Tur}
\altaffiliation[Present address: ]{Saint-Gobain Crystals, Hiram, OH 44234, USA}
\affiliation{National Superconducting Cyclotron Laboratory, Michigan State University, East Lansing, MI 48824-1321, USA}
\affiliation{Joint Institute for Nuclear Astrophysics, Michigan State University, East Lansing, MI 48824, USA}

\author{N. Verhanovitz}
\affiliation{National Superconducting Cyclotron Laboratory, Michigan State University, East Lansing, MI 48824-1321, USA}

\author{M. Yosoi}
\affiliation{Research Center for Nuclear Physics, Osaka University, Ibaraki, Osaka 567-0047, Japan}

\author{J. Yurkon}
\affiliation{National Superconducting Cyclotron Laboratory, Michigan State University, East Lansing, MI 48824-1321, USA}

\author{R.G.T. Zegers} \email{zegers@nscl.msu.edu}
\affiliation{National Superconducting Cyclotron Laboratory, Michigan State University, East Lansing, MI 48824-1321, USA}
\affiliation{Department of Physics and Astronomy, Michigan State University, East Lansing, MI 48824, USA}
\affiliation{Joint Institute for Nuclear Astrophysics, Michigan State University, East Lansing, MI 48824, USA}

\author{J. Zenihiro}
\affiliation{Research Center for Nuclear Physics, Osaka University, Ibaraki, Osaka 567-0047, Japan}

\date{\today}

\begin{abstract}
The $^{150}$Nd($^3$He,$t$) reaction at $140$ MeV/u and $^{150}$Sm($t$,$^3$He) reaction at $115$ MeV/u were measured, populating excited states in $^{150}$Pm. The transitions studied populate intermediate states of importance for the (neutrinoless) $\beta\beta$ decay of $^{150}$Nd to $^{150}$Sm. Monopole and dipole contributions to the measured excitation-energy spectra were extracted by using multipole decomposition analyses. The experimental results were
compared with theoretical calculations obtained within the framework of Quasiparticle Random-Phase Approximation (QRPA), which is one of the main methods employed for estimating the half-life of the neutrinoless $\beta\beta$ decay ($0\nu\beta\beta$) of $^{150}$Nd. The present results thus provide useful information on the neutrino responses for evaluating the $0\nu\beta\beta$ and $2\nu\beta\beta$ matrix elements.
The $2\nu\beta\beta$ matrix element calculated from the Gamow-Teller transitions through the lowest $1^{+}$ state in the intermediate nucleus is maximally about half of that deduced from the half-life measured in $2\nu\beta\beta$ direct counting experiments and at least several transitions through $1^{+}$ intermediate states in $^{150}$Pm are required to explain the $2\nu\beta\beta$ half-life.

Because Gamow-Teller transitions in the $^{150}$Sm($t$,$^3$He) experiment are strongly Pauli-blocked, the extraction of Gamow-Teller strengths was complicated by the excitation of the $2\hbar\omega$, $\Delta L=0$, $\Delta S=1$ isovector spin-flip giant monopole resonance (IVSGMR). However, the near absence of Gamow-Teller transition strength made it possible to cleanly identify this resonance, and the strength observed is consistent with the full exhaustion of the non-energy-weighted sum rule for the IVSGMR.
\end{abstract}


\maketitle

\section{Introduction}\label{sec:intro}
The two modes of $\beta\beta$ decay, 2-neutrino (2$\nu\beta\beta$) \cite{GMAY35} and 0-neutrino (0$\nu\beta\beta$) \cite{FUR39} have received a great deal of interest in the nuclear and particle physics communities.   A successful measurement of the 0$\nu\beta\beta$ decay half-life would confirm that neutrinos are Majorana rather than Dirac particles, could be used to constrain the neutrino mass, and would help solve the neutrino mass hierarchy \cite{AAL04,MOH07}.  Large-scale direct counting experiments have been launched to measure the $\beta\beta$ decay half-lives for several candidate nuclei (see e.g. Ref. \cite{AVI08}).

The 2$\nu\beta\beta$ decay half-life can be calculated with
(see e.g. Refs. \cite{AVI08,TOM91,MUT94,Ell02}):
\begin{equation} \label{eq:dbd_2nu}
[T^{2\nu}_{1/2}(0^+\rightarrow0^+)]^{-1}=G^{2\nu}(E_0,Z)|M^{2\nu}_{GT}|^2,
\end{equation}
where $G^{2\nu}(E_0,Z)$ is a phase-space factor, expressed in units of the electron mass squared, and M$^{2\nu}_{GT}$ is the double Gamow-Teller nuclear matrix element (NME). The latter is given by:
\begin{equation} \label{eq:dbd_GTME}
M^{2\nu}_{GT}=\displaystyle\sum_{j} \frac{\langle0^+_f\parallel\sum_{k}\sigma_{k}\tau^{-}_{k}\parallel1^+_j\rangle\langle1^+_j\parallel\sum_{k}\sigma_{k}\tau^{-}_{k}\parallel0^+_i\rangle}{E_j-E_0+Q_{\beta\beta}/2
},
\end{equation}
where $E_{j}-E_{0}$ is the energy difference between the ground state of the 2$\nu\beta\beta$ mother and the $j$-th $1^{+}$ state in the intermediate nucleus, and Q$_{\beta\beta}$ is the Q value for 2$\nu\beta\beta$ decay\footnote{Eq. (\ref{eq:dbd_GTME}) assumes the use of atomic masses in the calculation of Q$_{\beta\beta}$ and $E_{j}-E_{0}$.}.
The subscripts `$i$' and `$f$' refer to the 2$\nu\beta\beta$ mother and daughter states, respectively. The sum over $j$ refers to all states in the
intermediate nucleus, which are connected to the 2$\nu\beta\beta$ mother and daughter states via two ordinary $\beta$-decay GT matrix elements\footnote{Contributions from Fermi transitions are negligible, since the initial and final states are not members of the same isospin multiplet.}  $M_{j}(\textrm{GT}^{-})=\langle0^+_f\parallel\sum_{k}\sigma_{k}\tau^{-}_{k}\parallel1^+_j\rangle$ and $M_{j}(\textrm{GT}^{-})=\langle1^+_j\parallel\sum_{k}\sigma_{k}\tau^{-}_{k}\parallel0^+_i\rangle$. Here, the sum over $k$ runs over all the neutrons of the relevant decay nucleus. The latter matrix element can be accessed by ($p$,$n$)-type reactions on the initial nucleus and the conjugate of the first term $M_{j}(\textrm{GT}^{+})=\langle1^+_j\parallel\sum_{k}\sigma_{k}\tau^{+}_{k}\parallel0^+_f\rangle$ can be accessed by ($n$,$p$)-type reactions on the final nucleus.

The magnitude of these NMEs can be derived from Gamow-Teller transition strengths ($B(GT)$) via:
\begin{equation}\label{eq:nme}
|M_{j}(\textrm{GT}^{\pm})|^{2}= 
B_{j}(\textrm{GT}^{\pm}).
\end{equation}

The phases associated with each of the contributing transitions through the intermediate nucleus may interfere constructively or destructively. Therefore, theoretical methods are used to calculate the NME using constraints from experimental data. Abad \emph{et al.} \cite{ABA84} first hypothesized that the presence of a single low-lying state in the intermediate nucleus was sufficient to predict the 2$\nu\beta\beta$ decay half-life.  The validity of this Single-State Dominance (SSD) hypothesis has become a significant question.  It seems to apply to some $\beta\beta$-decay nuclei, although it is not clear whether transitions through higher-lying intermediate states do not contribute to the total matrix element or whether their contributions cancel \cite{SUH00a}.
More generally, it has been pointed out that $2\nu\beta\beta$ decay likely proceeds mainly through the low-lying Fermi-surface Single Quasi-Particle (FSQP) states in the intermediate nucleus \cite{EJI96,EJI09}.
Extracting the GT matrix elements of transitions to the intermediate states from data and applying Eq. (\ref{eq:dbd_GTME}) allows the SSD hypothesis and the role of FSQP and possible higher-lying states and resonances to be directly tested in comparison with experimentally-measured 2$\nu\beta\beta$ decay half-lives.

For the light Majorana neutrino exchange mechanism,
the half-life for 0$\nu\beta\beta$ decay can be calculated with
(see e.g. \cite{AVI08,TOM91,MUT94,Ell02}):
\begin{equation} \label{eq:dbd_0nu}
[T^{0\nu}_{1/2}(0^+\rightarrow0^+)]^{-1}=G^{0\nu}(E_0,Z)|M^{0\nu}|^2\langle m_{\nu} \rangle^2,
\end{equation}
where $G^{0\nu}(E_0,Z)$ is a phase-space factor and $m_{\nu}$ is the effective Majorana neutrino mass.  $M^{0\nu}$ is a sum over
over products of matrix elements of transitions to and from all the states in the intermediate nucleus.  In the case of 0$\nu\beta\beta$ decay the $\beta$ transitions occur at close range and are thus associated with large momentum transfers of $\sim 1$ fm$^{-1}$. Therefore,  transitions to intermediate states of many different spins and positive and negative parity can contribute to the matrix element. This greatly complicates the theoretical estimation of $T^{0\nu}_{1/2}(0^+\rightarrow0^+)$ and, as a consequence, closure approximations \cite{TOM91,MUT94} are sometimes applied.

The large phase-space factor $G^{0\nu}(E_0,Z)$ for $^{150}$Nd favors a shorter half-life of 0$\nu\beta\beta$ decay for this nucleus than those of the other $\beta\beta$-decaying nuclei
(see e.g. Refs. \cite{Ell02,EJI05}). In addition, the end-point energy ($Q_{\beta\beta}=3.37$ MeV) is high, which is preferred for counting experiments since background contributions are reduced. $^{150}$Nd is, therefore, considered to be one of the most promising candidates for experimental searches of 0$\nu\beta\beta$ decay, and may be the focus of three planned experiments: SNO+ \cite{KRA10}, SuperNEMO \cite{DEP10}, and DCBA \cite{ISH96,ISH08}. However, $^{150}$Nd and its $\beta\beta$ decay daughter $^{150}$Sm are both deformed nuclei and the difference in deformation between the two is expected to reduce the matrix elements and thus increase the half-life, partially mitigating the effect of the high phase-space factor
\cite{FAN10a,FAN11}.
Nevertheless, based on the recent calculations that include the effects of deformation, Fang $\emph{et al.}$ still conclude that the 0$\nu\beta\beta$ decay of $^{150}$Nd may provide one of the best probes of the Majorana neutrino mass \cite{FAN10a,FAN11}.
A similar conclusion was drawn in the calculations employing
the interacting boson model~\cite{Bar09}, and the generator-coordinate method with particle number and angular momentum projection~\cite{Rodri10}.

As one of the heaviest $\beta\beta$ emitters, efforts to model the transition between $^{150}$Nd and $^{150}$Sm are hindered by the complexity of the nuclear systems and by the lack of experimental information on the intermediate nucleus, $^{150}$Pm. Dvornick\'y \emph{et al.} \cite{DVO07} postulated that fulfillment of the SSD hypothesis for 2$\nu\beta\beta$ in $^{150}$Nd is not expected unless an unknown low-lying 1$^+$ state in $^{150}$Pm is found experimentally. At present, the Evaluated Nuclear Structure Data File (ENSDF) only lists the half-life and a tentative J$^{\pi}=1^{-}$ assignment for the ground state of $^{150}$Pm \cite{DER95}.

\begin{figure}
\centering
\includegraphics[]{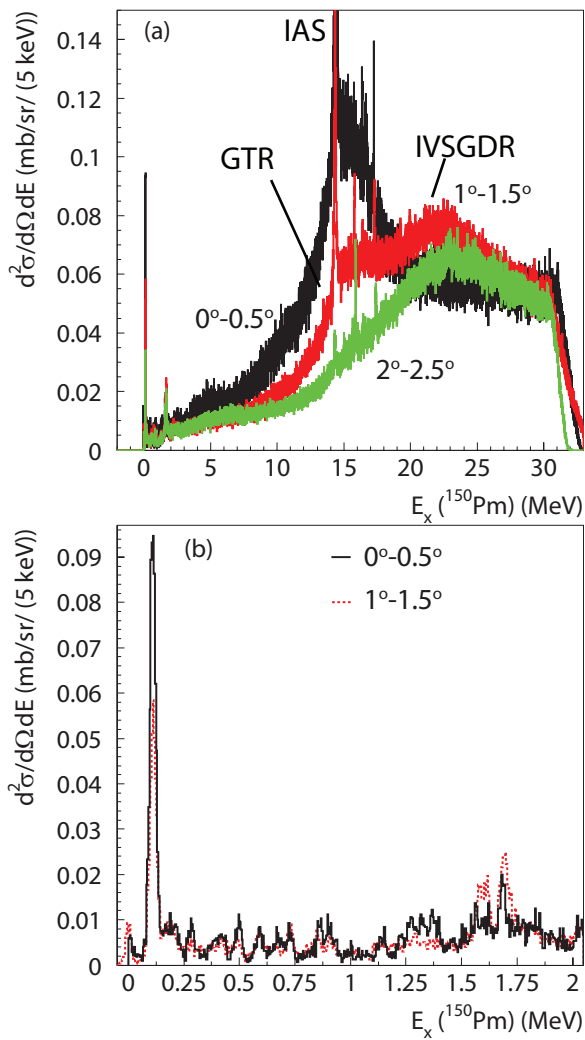}
\caption{(color online) (a) Excitation-energy spectra measured in the $^{150}$Nd($^3$He,$t$) experiment at $E(^{3}$He)=140 MeV/u. Spectra for scattering-angle bins from $0^{\circ}$--$0.5^{\circ}$ (black), $1^{\circ}$--$1.5^{\circ}$ (red) and $2^{\circ}$--$2.5^{\circ}$ (green) are shown. Excitations of the IAS, GTR, and IVSGDR are indicated in the figure. Note that the excitation of the IAS exceeds the y-axis scale for the first two angular bins shown.  The IAS and GTR are strongly forward-peaked, while the IVSGDR peaks at $\sim 1.5^ \circ$. Peaks from the $^{16}$O($^{3}$He,$t$) reaction at 16 and 17.5 MeV indicate some oxidation of the target surface had taken place.  b) Expanded excitation-energy spectra below $E_{x}(^{150}$Pm)=2 MeV and scattering angles of $0^{\circ}$--$0.5^{\circ}$ (black) and $1^{\circ}$--$1.5^{\circ}$ (dashed-red) are shown.}
\label{Nd_cs1}
\end{figure}

The experimental nuclear physics community has made a concerted effort to provide constraints on the theoretical calculations for both 2$\nu$- and 0$\nu$-$\beta\beta$ decay using several complementary techniques, including high-precision Q$_{\beta\beta}$ measurements \cite{KOH10,RED09,MOU10,RAH09},  the determination of valence proton and neutron orbits using single-nucleon transfer \cite{SCH08,KAY09}, and direct population of states in the intermediate nucleus via charge-exchange experiments \cite{DOH08,YAK09,GRE08a,GRE08,AKI97,RAK04,GRE07,EJI00}.  Charge-exchange reactions can directly populate states in the intermediate nucleus, and this makes them a valuable tool to help constrain calculations of nuclear matrix elements.

In this work, the results of two charge-exchange experiments are presented. These experiments aimed to shed light on the nuclear structure relevant for studies of $\beta\beta$ decay of $^{150}$Nd. In the first experiment, the $^{150}$Nd($^{3}$He,$t$) reaction at 140 MeV/u was studied to extract information about the first leg of the $\beta \beta$ transition from the $^{150}$Nd mother to $^{150}$Pm. In the second experiment, the $^{150}$Sm($t$,$^{3}$He) reaction at 115 MeV/u was investigated to acquire information about the second leg of the $\beta\beta$ transition from $^{150}$Pm to the $^{150}$Sm daughter. In both experiments, Gamow-Teller (GT; associated with angular momentum transfer $\Delta L=0$, and spin transfer $\Delta S=1$) and isovector spin-dipole ($\Delta L=1$, $\Delta S=1$) strength distributions were extracted and compared with theoretical calculations in the framework of Quasiparticle Random-Phase Approximation (QRPA) that are also used for calculations of $0\nu\beta\beta$ and $2\nu\beta\beta$ (GT strengths only) decay \cite{YOU09,FAN10,FAN10a,FAN11}. The comparison serves as a test of the QRPA calculations and can be used to further improve this and future theoretical work on the $\beta \beta$ decay of $^{150}$Nd. In addition, by combining the results from the two experiments, the possibility of SSD for the $2\nu\beta\beta$ of $^{150}$Nd was investigated.


Charge-exchange (CE) reactions are characterized by the transfer of one unit of isospin ($\Delta T=1$).  At forward scattering angles, transitions associated with small angular momentum transfer (i.e. monopole ($\Delta L=0$) and dipole ($\Delta L=1$)) are preferably excited. Popular choices of probes include the (p,n) and ($^3$He,$t$) reactions in the $\Delta T_z=-1$ direction and the ($n$,$p$), ($d$,$^2$He), and ($t$,$^3$He) reactions in the $\Delta T_z=+1$ direction, although pionic and heavy-ion probes have also been used \cite{HAR01}. For probes not intrinsically selective to a specific spin-transfer, excitations associated with spin-transfer ($\Delta$S=1) are nevertheless strongly favored over those without spin-transfer ($\Delta S=0$) for beam energies in excess of $\gtrsim$100 MeV/u \cite{LOV81,FRA85,RAP94}. At such beam energies, multi-step contributions are small and the CE reaction can be considered a direct, single-step process.

\begin{figure}
    \centering
    \includegraphics[scale=1.0]{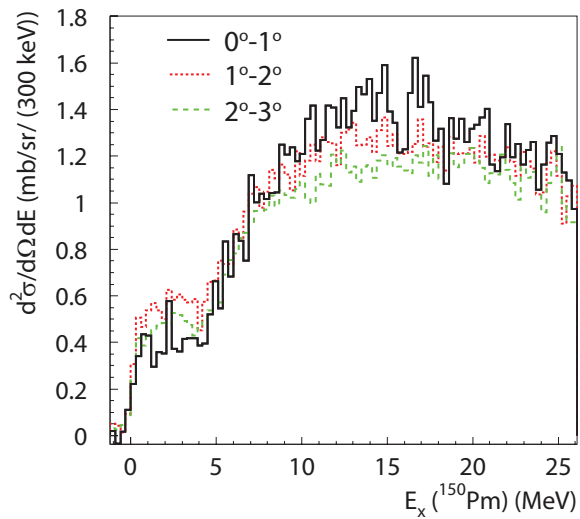}
    \caption[]{(color online) Differential cross sections for $^{150}$Sm($t$,$^3$He) reaction at $E(t)$=115 MeV/u over the full excitation energy covered in the experiment.  Data are grouped into 1$^\circ$-wide angular bins of which the first 3 out of five are shown in the figure: 0$^\circ$--1$^\circ$ (black, solid line), 1$^\circ$--2$^\circ$ (red, dotted line) and 2$^\circ$--3$^\circ$ (green, dashed line).}
    \label{Smcs}
\end{figure}

For GT transitions, and in the limit of vanishing linear momentum transfer $q$, the charge-exchange differential cross section is proportional to GT strength \cite{TAD87}:
\begin{equation}\label{eq:CD_propeq2}
\frac{d\sigma}{d\Omega}\Bigl|_{q=0} = \hat{\sigma}_{GT}B(GT),
\end{equation}
where $\hat{\sigma}_{GT}$ is the unit cross section.  A similar proportionality between differential cross sections and transition strength has been established for Fermi transitions, but there is no proven equivalent for dipole or higher multipole transitions.  The GT unit cross section for the ($^3$He,$t$) and ($t$,$^3$He) reactions can be found using the phenomenologically established relationship \cite{ZEG07,PER11}:
\begin{equation}
\hat{\sigma}_{GT}=109/A^{0.65}.
\label{eq:ugt}
\end{equation}
For Fermi transitions ($\Delta$L=0, $\Delta$S=0) (confined to the excitation of the isobaric analog state (IAS) in the ($^{3}$He,$t$) reaction), the unit cross section is reasonably well described by \cite{ZEG07}:
\begin{equation}
\hat{\sigma}_{F}=72/A^{1.06}.
\label{eq:uf}
\end{equation}

CE reactions are an excellent tool for studying isovector giant resonances, which can be described macroscopically as out-of-phase density oscillations of the proton and neutron fluids in the nucleus, or microscopically as coherent superpositions of one-particle, one-hole (1$p$--1$h$) excitations \cite{HAR01}. Of the giant resonances associated with spin-transfer, the Gamow-Teller Resonance (GTR) and Isovector Spin-flip Giant Dipole Resonance (IVSGDR) are most-widely studied (see Ref. \cite{HAR01} and references therein). Experimental information about the Isovector Spin-flip Giant Monopole Resonance (IVSGMR) is more scant, however. This resonance, like the GTR, is associated with $\Delta$L=0, $\Delta$S=1, but microscopically described by $2\hbar\omega$ (i.e., over two major shells) $1p$-$1h$ excitations. Tentatively observed in ($^{3}$He,$t$) experiments at 200 MeV/u and 300 MeV/u \cite{ELL83,AUE83} and a comparative study of ($\vec{p},\vec{n}$) reactions at 200 and 800 MeV, its existence was confirmed in $^{208}$Pb($^{3}$He,$tp$) experiments at $59$ MeV/u \cite{ZEG00,ZEG01} and $137$ MeV/u \cite{ZEG03}. In the $\Delta$T$_{z}=-1$ (i.e., ($p$,$n$)) direction, the IVSGMR is located at relatively high excitation energies ($\sim35$ MeV) \cite{AUE84}. The presence of the continuum, in combination with the large width of the IVSGMR ($\Gamma\sim 10$ MeV) makes it is difficult to study this resonance experimentally. In the $\Delta$T$_{z}=+1$ (i.e. ($n$,$p$)) direction, the IVSGMR is expected to be situated at lower excitation energies ($\sim 15-20$~MeV) \cite{AUE84}, which should make observation easier. Indeed, in a $^{58}$Ni($t$,$^{3}$He) experiment at $43$ MeV/u \cite{GUI06}, evidence for significant, albeit somewhat fragmented, $\Delta L=0$ strength was found that could be associated with the excitation of the IVSGMR. Because of the relatively low beam energy used in that experiment, its non-spin-flip companion, the Isovector Giant Monopole Resonance (IVGMR), also contributed significantly to the $\Delta L=0$ response.

A complication in the search for the IVSGMR is that both the GTR and IVSGMR are associated with $\Delta L=0$ and their differential cross sections thus have similar angular distributions. High-lying GT strength can, therefore, not easily be distinguished from strength due to the excitation of the IVSGMR (see e.g. Ref. \cite{YAK05}) in a single experiment\footnote{It is noted that the transition density of the IVSGMR has a node near the surface, in contrast to that of the GTR. Therefore, it is possible to extract information about the separate contributions from the two excitations by comparing the monopole strength distributions extracted from an experiment utilizing a reaction that probes the surface of the nucleus (the ($^{3}$He,$t$), ($t$,$^{3}$He) or heavy-ion reactions), with an experiment employing a reaction that probes the interior more strongly (e.g. the ($p$,$n$) or ($n$,$p$) reactions) \cite{AUE89,AUE98}.}.
However, in medium-heavy nuclei where GT excitations are strongly suppressed in the $\Delta$T$_{z}=+1$ direction due to Pauli-blocking effects, the IVSGMR can dominate the $\Delta$L=0 response. If  $^{150}$Sm were spherical, its 62 protons would fill a fraction of $4\hbar\omega$ ($3s$, $2d$ and $1g$) orbits. The 88 neutrons completely fill the $4\hbar\omega$ orbits, thereby completely blocking $0\hbar\omega$ GT excitations in the $\Delta$T$_{z}=+1$ direction. Given the relatively large deformation of $^{150}$Sm ($\beta=0.19$ \cite{FAN10a}), this picture is too simple: the proton $1h_{11/2}$ orbit is likely partially filled, allowing for proton-$1h_{11/2}$, neutron-$1h_{9/2}$ Gamow-Teller amplitudes  associated with transitions to low-lying $1^{+}$ states in $^{150}$Pm. Nevertheless, given the strong Pauli blocking and the near absence of GT strength, the study of the $^{150}$Sm($t$,$^{3}$He) reaction makes it easier to identify the IVSGMR. Finding evidence for the excitation of this resonance was a secondary goal of the $^{150}$Sm($t$,$^{3}$He) experiment.

After giving an overview of the experimental setups and procedures for the $^{150}$Nd($^{3}$He,$t$) experiment at $140$ MeV/u and $^{150}$Sm($t$,$^{3}$He) experiment at $115$ MeV/u in Section \ref{sec:exp}, the extraction of $\Delta L=0$, $\Delta S=1$ (GT and IVSGMR) and $\Delta L=1$, $\Delta S=1$  (spin-dipole) contributions to the total response for both reactions is discussed in Section \ref{sec:gt}. The comparison between the experimental results and theoretical calculations is covered in Section \ref{sec:comp}. A discussion of implications of the extracted GT strengths for $2\nu\beta\beta$ decay and the applicability of SSD is presented in Section \ref{sec:me}.

\begin{figure}
  \centering
    \includegraphics[scale=1.0]{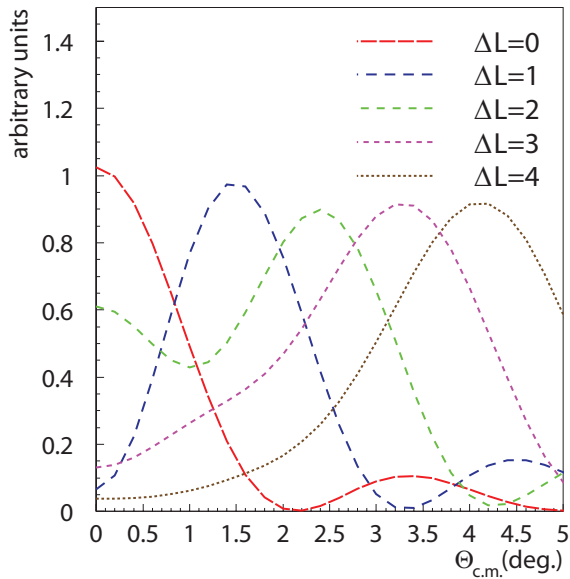}
    \caption[]{(color online) Calculated angular distributions for transitions with $\Delta$L=0-4 via the $^{150}$Nd($^{3}$He,$t$) reaction at 140 MeV/u. The relative scaling of the displayed cross sections is arbitrary and the calculations were performed at $E_{x}=0$.}
    \label{Nd_shapes}
\end{figure}

\section{Experimental Setup and Procedures}\label{sec:exp}
\subsection{The $^{150}$Nd($^3$He,$t$) experiment}\label{sec:ndexp}
The $^{150}$Nd($^3$He,$t$)$^{150}$Pm* experiment took place at the Research Center for Nuclear Physics (RCNP).  A $^3$He$^{2+}$ beam was accelerated to $140$ MeV/u in the coupled AVF and Ring cyclotrons.  The $^3$He$^{2+}$ beam, with an intensity of up to $5\times10^{10}$ s$^{-1}$, was impinged on a  1.0-mg/cm$^2$ thick metallic $^{150}$Nd foil with an isotopic purity of 96\%. Tritons from the ($^3$He,$t$) reaction  were detected in the focal plane of the Grand Raiden spectrometer \cite{FUJ99}. Data were acquired at scattering angles of 0$^\circ$, 2.5$^\circ$, and 4$^\circ$. Each setting covered an angular range of about 2$^{\circ}$ and differential cross sections in the angular range between 0$^\circ$ and 5$^\circ$ were extracted.

The detector array in the focal plane of the Grand Raiden spectrometer consisted of two sets of multi-wire drift chambers (MWDCs) for position and angle measurements and two 10-mm thick plastic scintillators for energy-loss measurements.  A hit in the first scintillator also served as the event trigger and the start of a time-of-flight (TOF) measurement.  The stop signal was produced by the cyclotron RF signal. By combining the energy-loss signal and the TOF information, tritons were uniquely identified.

The beam spot was momentum-dispersed on the target to match the dispersion of the spectrometer in order to optimize the momentum resolution \cite{FUJ02}. The ion-optics of the spectrometer was tuned to run in over-focus mode \cite{FUJ01} to simultaneously achieve good angular resolutions in the dispersive and non-dispersive planes.  A calibration measurement using a sieve-slit was used for the determination of the parameters of a ray-trace matrix for reconstructing scattering angles at the target from position and angle measurements in the focal plane (for details, see e.g. Ref. \cite{ZEG04}). Triton energies were calibrated from $^{nat}$Mg($^3$He,$t$) reactions to known states in $^{24,25,26}$Al.  The excitation-energy resolution was 33 keV at full-width-half-maximum (FWHM), and the resolution of the reconstructed laboratory scattering angle was 0.42$^\circ$ (FWHM).

Beam intensities were monitored by integrating the charge of the $^3$He$^{2+}$ beam in Faraday cups. The cups used at the three different angular settings were cross-calibrated by measuring elastic scattering rates of $^{3}$He beam particles on hydrogen (in a CH$_{2}$ target) in the beam line from the cyclotrons to the spectrometer. The cross section extracted for the $^{26}$Mg($0^{+}_{g.s.}$)($^{3}$He,$t$)$^{26}$Al($1^{+}$, 1.06 MeV) reaction using the $^{nat}$Mg target was used to confirm the consistency with cross sections reported in Refs. \cite{ZEG06,ZEG07}. Although this procedure does not completely rule out a common systematic error in the determination of absolute cross sections in this and earlier experiments, it allows one to employ the phenomenologically extracted mass-dependent equation for the GT and Fermi unit cross section of Ref. \cite{ZEG07,PER11}.

In addition to the above measurements, $^{150}$Nd($^3$He,$^3$He$^{\prime}$) elastic scattering data were taken between $8^{\circ}$ and $20^{\circ}$ degrees in the center-of-mass frame.  The differential cross sections were fitted with the code ECIS97 \cite{ECIS} to extract optical model parameters that serve as input for Distorted-Wave Born Approximation (DWBA) calculations for the $^{150}$Nd($^{3}$He,$t$) reaction (discussed further in Section \ref{sec:gt}). The optical model contained real and imaginary volume Woods-Saxon potentials. The fitted parameters were $-58.57$ MeV, 1.134 fm and 1.032 fm for the depth ($V$), radius parameter ($r_{v}$) and diffusiveness ($a_{v}$) of the real Woods-Saxon potential and $-66.70$ MeV, 1.093 fm, and 0.94 fm for the depth ($W$), radius parameter ($r_{w}$) and diffusiveness ($a_{w}$) of the imaginary Woods-Saxon potential.

\begin{figure*}
  \centering
    \includegraphics[scale=0.75]{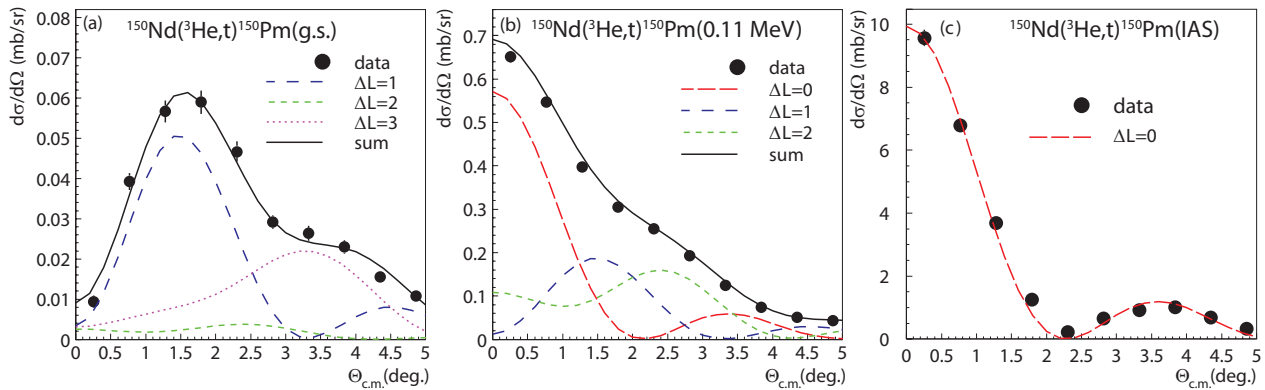}
    \caption[]{(color online) Differential cross sections for the excitation of (a) the ground state, (b) the strong state observed at 0.11 MeV and (c) the IAS at 14.35 MeV in $^{150}$Pm via the $^{150}$Nd($^{3}$He,$t$) reaction at $140$ MeV/u. A MDA is performed for the first two, whereas the excitation of the IAS is well reproduced by a pure $\Delta L=0$ angular distribution. Error bars are smaller than the data markers for cases where the bars are not visible. See text for details.}
    \label{Nd_low2}
\end{figure*}

The angular range covered  in the $^{150}$Nd($^3$He,$t$) experiment was divided into ten 0.5$^\circ$-wide bins.  Fig. \ref{Nd_cs1}(a) shows differential cross sections for three such bins up to an excitation energy of 30 MeV in $^{150}$Pm. The spectrum shape changes as a function of scattering angle due to the presence of states and giant resonances associated with different angular momentum transfer.  The angular distributions of the IAS at 14.35 MeV and GTR (centered at $\sim 15.25$ MeV, but with tails extending down to 5 MeV and up to 20 MeV) peak at 0$^\circ$, while that of the IVSGDR (centered at $\sim 22.8$ MeV) peaks around 1.5$^\circ$.  At larger scattering angles, the spectrum becomes more featureless. Narrow states seen at excitation energies above the IAS are due to $^{16}$O($^{3}$He,$t$) reactions from minor oxygen contamination of the target and which appear due to the high resolution achieved in the experiment.
At low excitation energies, several discrete states are observed and the region up to 2 MeV is expanded in Fig. \ref{Nd_cs1}(b). The spectra for two angular bins are shown: $0^{\circ}$--$0.5^{\circ}$, where GT excitations peak and $1.0^{\circ}$--$1.5^{\circ}$, where dipole excitations peak. The strongest transition is to a state at 0.11 MeV, which clearly peaks at forward scattering angles. Several other GT transitions to discrete states appear up to 1.5 MeV. The transition to the ground state of $^{150}$Pm appears to be associated with a dipole transition, and several other dipole transitions populate discrete states around 1.6 MeV. We note that the high level density makes analysis of individual states challenging even at low excitation energies, and impossible above $\sim$ 2 MeV.
The detailed analysis of the $^{150}$Nd($^3$He,$t$) data is discussed in Section \ref{sec:gt}.

\begin{figure}
  \centering
    \includegraphics[width=0.5\textwidth]{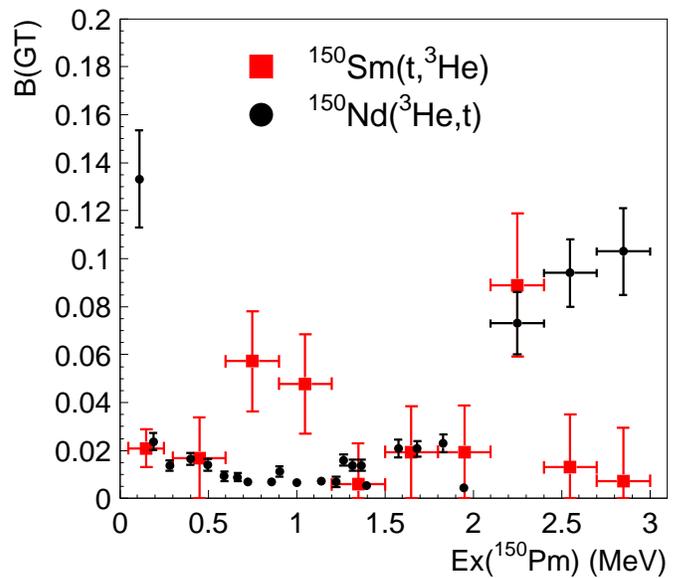}
    \caption[]{(color online) Extracted GT strengths from the two experiments in this work, for excitation energies between 0 MeV and 3 MeV in $^{150}$Pm.  Note that the $^{150}$Nd data refers to the excitation of individual states, except for $E_{x}=2-3$ MeV, where three data point cover the region between 2.1 and 3.0 MeV. The $^{150}$Sm data refer to strengths in 300-keV wide bins, except for the first data point, for which the location of the strength could be limited to the region between 50 and 250 keV. }
    \label{dis_Bgt}
\end{figure}

\subsection{The $^{150}$Sm($t$,$^3$He) experiment at NSCL}\label{subsec:smexp}
The $^{150}$Sm($t$,$^3$He) experiment was performed at the National Superconducting Cyclotron Laboratory in a campaign that also included a measurement of the $^{13}$C($t$,$^{3}$He) reaction using the same setup and which was discussed in Ref. \cite{GUE09}.

A primary beam of $^{16}$O was accelerated to 150 MeV/u in the coupled K500 and K1200 cyclotrons \cite{CCF88} and impinged upon a 3526 mg/cm$^2$ $^{nat}$Be production target.  The fragments were fed into the A1900 fragment separator \cite{MOR03}, producing a $115$ MeV/u secondary triton beam \cite{HIT06}.  The momentum spread of the triton beam was limited to $\frac{dp}{p} = \pm 0.25\%$ using slits in the fragment separator. The tritons were then transported to the reaction target placed at the pivot point of the S800 spectrometer \cite{BAZ03}.  $^{16}$O primary beam intensities were monitored with non-intercepting probes placed at the exit of the K1200 cyclotron and calibrated against an absolute measurement of the triton-beam intensity at the reaction  target measured using a plastic scintillator. Readings from the non-intercepting probes were then used throughout the experiment to determine the integrated triton beam on target.
The transmission from the A1900 to the reaction target was 85\%, where a triton-beam intensity of 10$^7$ s$^{-1}$ was achieved.

As in the ($^{3}$He,$t$) experiment, the triton beam was dispersed over the reaction target to match the dispersion of the spectrometer and maximize the momentum resolution. Given the relatively large momentum spread of the secondary beam and the momentum dispersion of 10.5 cm/\% of the spectrometer, the beam-spot size was approximately 5-cm wide in the dispersive direction and 1-cm wide in the non-dispersive direction. Therefore, a relatively large $^{150}$Sm target (7.5 cm by 2.5 cm) was used, with an isotopic purity of 96\%. Given the relatively low beam intensity compared to stable beam experiments, the target was produced with a thickness of 18.0 mg/cm$^2$.

The production of this $^{150}$Sm target was a challenge. The available isotopically pure $^{150}$Sm material could not be rolled at the thickness and size required. Therefore, a method to efficiently evaporate the samarium  with electron-beam technology was developed.  To achieve the desired area and uniformity of thickness, the samarium was evaporated onto two slides of 7.5 cm by 2.5 cm reaching a thickness of 9.0 mg/cm$^2$ each.
The two foils were then stacked to achieve a thickness of 18.0 mg/cm$^2$.  A total amount of 1.5 grams of $^{150}$Sm was used in the evaporation. The difference in energy loss between tritons and $^{3}$He particles resulted in an ambiguity of reconstructed excitation energy in $^{150}$Pm of about 200 keV. In addition to the $^{150}$Sm target, two calibration targets were used: 10-mg/cm$^2$ thick $^{12}$CH$_2$ and 18-mg/cm$^2$ thick $^{13}$CH$_2$ \cite{GUE09}.

\begin{figure}
  \centering
    \includegraphics[scale=1.0]{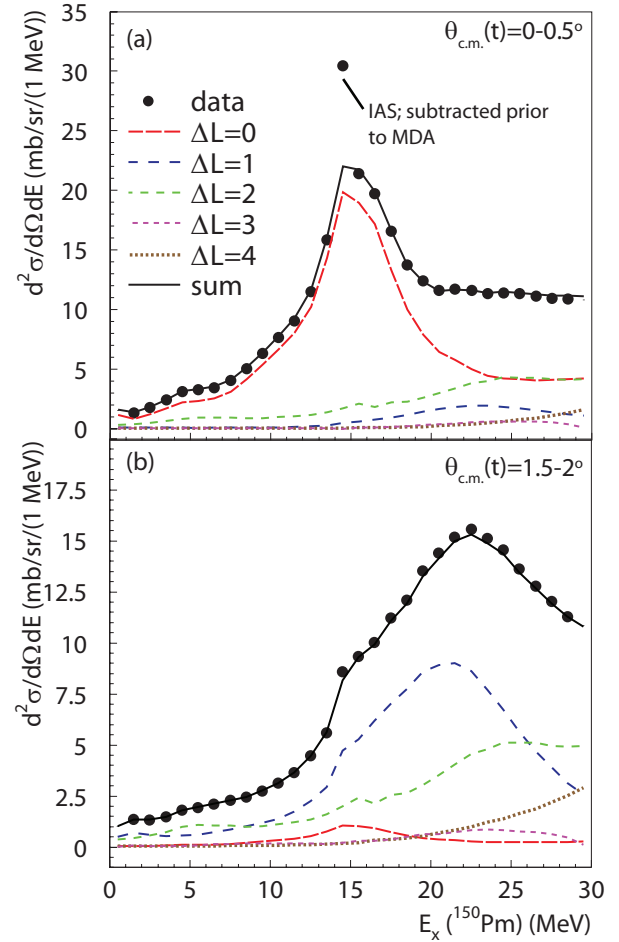}
\caption[]{(color online) Results of the MDA of the $^{150}$Nd($^3$He,$t$) data. Shown are the decompositions of the spectra at 0$^\circ$-0.5$^\circ$ (a) and 1.5$^\circ$-2$^\circ$ (b). The peak due to the IAS was subtracted from the data prior to the MDA and thus does not appear in the decomposed results. The GTR dominates the spectrum at 0$^\circ$-0.5$^\circ$ and the IVSGDR dominates at 1.5$^\circ$-2$^\circ$. In both (a) and (b), the results in each 1-MeV wide bin from the MDA for each multipole are connected by lines, rather then showing the individual points.}
\label{Nd_md_summary}
\end{figure}

$^3$He ejectiles were momentum analyzed in the S800 spectrometer \cite{YUR99,BAZ03}. Tracking in the focal plane was performed with two cathode-readout drift chambers (CRDCs).
A dual scintillator stack recorded energy-loss information and the first scintillator also provided the event trigger and the start of a TOF measurement, with the stop provided by the cyclotron RF signal. The energy-loss signal and the timing information were used to uniquely identify the $^{3}$He particles.

A fifth-order raytrace matrix, calculated with the code \textsc{COSY Infinity} \cite{COSY}, was used to reconstruct the scattering angles in the dispersive and non-dispersive planes, the hit position of the beam at the reaction target in the non-dispersive plane, and the energy of the $^{3}$He particles.  The excitation energy of $^{150}$Pm was reconstructed using a missing-mass calculation. A resolution of 300 keV (FWHM) was achieved. The systematic error in the excitation energy was estimated to be maximally 50 keV, based on data from the $^{13}$C($t$,$^3$He) reaction \cite{GUE09}. Non-dispersive and dispersive angles at the target were used to reconstruct the scattering angle. The angular resolution was 0.6$^\circ$ (FWHM).
The S800 spectrometer was set at $0^{\circ}$, and differential cross sections could be measured from 0$^\circ$ to 5$^\circ$.  The S800 acceptance, which is a function of the non-dispersive hit position of the beam on the target, the momentum of the $^{3}$He particles and the non-dispersive and dispersive components of the scattering angle, was modeled in a Monte-Carlo simulation.

In a previous experiment on a $^{64}$Zn target \cite{HIT09}, a minor amount of background was observed in the ($t$,$^3$He) spectrum that could relatively easily be subtracted. This background appeared stronger in the runs with the $^{150}$Sm target. Further investigations during the experiment presented here indicated that these were due to the $^6$He $\rightarrow$ $^3$He + ${3n}$ breakup reactions from a small amount of $^{6}$He ($\sim$15\%) present in the secondary beam.  About halfway through the $^{150}$Sm($t$,$^3$He) experiment, a 195 mg/cm$^2$ wedge was inserted at the intermediate image of the A1900 spectrometer, removing the $^{6}$He contaminant from the beam and the background from the measured excitation energy spectra. The background in the earlier part of the experiment could be characterized by comparing spectra taken before and after insertion of the wedge. Its relatively flat momentum and angular distributions made it easy to subtract from the data.

A small amount of hydrogen was absorbed on to the $^{150}$Sm target. However, the cross section for the $^{1}$H($t$,$^{3}$He) reaction is very large and it was visible in the excitation-energy spectra. The difference in $Q$-value between the $^{1}$H($t$,$^{3}$He) and $^{150}$Sm($t$,$^3$He)$^{150}$Pm(g.s.) reactions is only 2.67 MeV. Since the recoil energy increases much more rapidly with increasing scattering angle for the former than for the latter reaction, events from the $^{1}$H($t$,$^{3}$He) reaction contaminated the $^{150}$Pm excitation energy spectrum at scattering angles beyond $2^{\circ}$.  Data taken with a $^{nat}$CH$_2$ target were used to model and subtract the hydrogen contribution to the $^{150}$Sm($t$,$^3$He) excitation energy spectra. Statistical and systematic errors due to the subtraction of reactions from the hydrogen contaminant and from the breakup of $^{6}$He in the first half of the experiment were taken into account in the subsequent analysis of the data.

\begin{table}
\caption[]{\label{tb:Nd_bgt}GT strengths extracted from the $^{150}$Nd($^3$He,$t$) experiment. Values on the left side  provide the strengths for 1-MeV wide bins in excitation energies, whereas the values on the right side are for individual peaks observed at excitation energies below 2.1 MeV and for the three 300-keV wide bins between 2.1 and 3.0 MeV.}
\begin{ruledtabular}
\begin{tabular}{cccc}
\multicolumn{2}{c}{1 MeV bins} & \multicolumn{2}{c}{$E_{x}$($^{150}$Pm$)<3.0$ MeV} \\ \cline{1-2} \cline{3-4}
\noalign{\smallskip}
E$_x$($^{150}$Pm)& B(GT) & E$_x$($^{150}$Pm)\footnotemark[1] & B(GT) \\
(MeV) & &  (MeV) & \\ \cline{1-2} \cline{3-4}
\noalign{\smallskip}
0--1 & 0.27 $\pm$ 0.04 &  0.11 & 0.133 $\pm$ 0.020 \\
1--2 & 0.19 $\pm$ 0.03 &  0.19 & 0.023 $\pm$ 0.004 \\
2--3 & 0.27 $\pm$ 0.04 &  0.28 & 0.013 $\pm$ 0.002 \\
3--4 & 0.38 $\pm$ 0.06 &  0.40 & 0.016 $\pm$ 0.003 \\
4--5 & 0.49 $\pm$ 0.07 &  0.50 & 0.013 $\pm$ 0.003 \\
5--6 & 0.51 $\pm$ 0.08 &  0.59 & 0.009 $\pm$ 0.002 \\
6--7 & 0.55 $\pm$ 0.08 &  0.67 & 0.009 $\pm$ 0.002 \\
7--8 & 0.68 $\pm$ 0.10 &  0.73 & 0.007 $\pm$ 0.002 \\
8--9 & 0.90 $\pm$ 0.14 &  0.86 & 0.007 $\pm$ 0.002 \\
9--10 & 1.17 $\pm$ 0.18 &  0.90 & 0.011 $\pm$ 0.002 \\
10--11 & 1.46 $\pm$ 0.22 &  1.00 & 0.006 $\pm$ 0.002 \\
11--12 & 1.77 $\pm$ 0.27 &  1.14 & 0.007 $\pm$ 0.002 \\
12--13 & 2.28 $\pm$ 0.34 &  1.23 & 0.006 $\pm$ 0.002 \\
13--14 & 3.24 $\pm$ 0.49 &  1.27 & 0.015 $\pm$ 0.002 \\
14--15 & 4.55 $\pm$ 0.70 &  1.32 & 0.013 $\pm$ 0.002 \\
15--16 & 4.43 $\pm$ 0.67 &  1.37 & 0.013 $\pm$ 0.002 \\
16--17 & 4.10 $\pm$ 0.62 &  1.40 & 0.005 $\pm$ 0.001 \\
17--18 & 3.25 $\pm$ 0.49 &  1.58\footnotemark[2] & 0.020 $\pm$ 0.004 \\
18--19 & 2.52 $\pm$ 0.38 &  1.68\footnotemark[2] & 0.019 $\pm$ 0.003 \\
19--20 & 2.05 $\pm$ 0.31 &  1.83 & 0.022 $\pm$ 0.004 \\
20--21 & 1.72 $\pm$ 0.26 &  1.95 & 0.004 $\pm$ 0.001 \\ \cline{3-4}
\noalign{\smallskip}
21--22 & 1.60 $\pm$ 0.24 &  $\sum_{\textrm{0-2 MeV}}$ & 0.37$\pm$0.03$\pm$0.04\footnotemark[3]\\
22--23 & 1.44 $\pm$ 0.22 &  & \\
23--24 & 1.33 $\pm$ 0.20 &  2.1--2.4 & 0.073$\pm$0.013\footnotemark[4]\\
24--25 & 1.33 $\pm$ 0.20 &  2.4--2.7 & 0.094$\pm$0.014\footnotemark[4]\\
25--26 & 1.36 $\pm$ 0.21 &  2.7--3.0 & 0.103$\pm$0.018\footnotemark[4]\\ \cline{3-4}
\noalign{\smallskip}
26--27 & 1.39 $\pm$ 0.21 &  $\sum_{\textrm{0-3 MeV}}$ & 0.64$\pm$0.04$\pm$0.06\footnotemark[3]\\
27--28 & 1.49 $\pm$ 0.22 &  & \\
28--29 & 1.59 $\pm$ 0.24 &  & \\
29--30 & 1.71 $\pm$ 0.26 &  & \\
\cline{1-2}
\noalign{\smallskip}
$\sum_{\textrm{0-30 MeV}}$ & 50.0$\pm$1.7$\pm$5.0\footnotemark[3]\\
\end{tabular}
\end{ruledtabular}
\footnotetext[1]{The uncertainty in the excitation energy for each level is 10 keV.}
\footnotetext[2]{These levels are predominantly associated with $\Delta L=1$ and the GT component is only a minor contribution. The excitation energies associated with the GT components are relatively uncertain (20 keV compared to 10 keV for the other states).}
\footnotetext[3]{The first error represents statistical and systematical errors that were uncorrelated from one peak/energy bin to another. The second error represents the uncertainty in the unit cross section, which is correlated for all peaks/bins (see text for more details).}
\footnotetext[4]{These GT strength are for the full 300-keV wide energy bin and could not be associated with transitions to particular states.}
\end{table}

\begin{figure}
  \centering
    \includegraphics[scale=1.0]{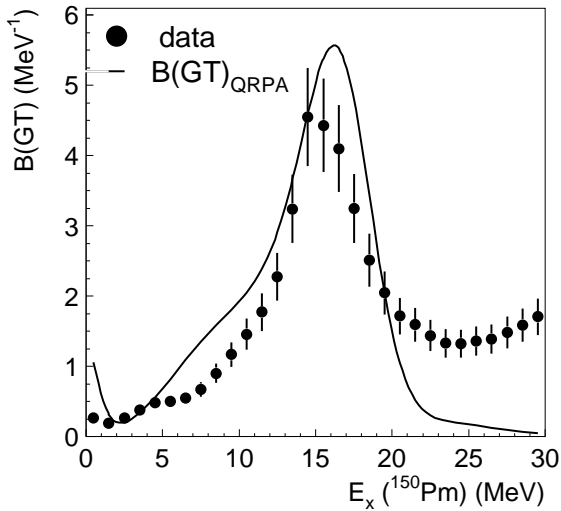}
\caption[]{GT strengths extracted from the $^{150}$Nd($^3$He,$t$) experiment and the comparison with calculated values in QRPA (solid line).}
\label{Nd_gt_th}
\end{figure}

The angular range of 0$^\circ$-5$^\circ$ was subdivided into five $1^{\circ}$-wide bins. The $^{150}$Pm excitation-energy spectra for the first 3 angular bins are shown in Fig. \ref{Smcs}.  Compared to the spectra from the $^{150}$Nd($^3$He,$t$) experiment, the spectra are rather featureless. This is partially due to the fact that transitions to individual final states are not easily discernable because the energy resolution is worse. In addition, there is no IAS, and the GTR is strongly Pauli blocked (as discussed in Section \ref{sec:intro}). For $E_{x}(^{150}$Pm)=1--5 MeV, the cross sections peak between $1^{\circ}$ and $2^{\circ}$, indicating contributions from dipole excitations. For $E_{x}(^{150}$Pm)=8--20 MeV, excess cross section is observed at forward scattering angles, suggestive of monopole contributions. A detailed analysis of the $^{150}$Sm($t$,$^3$He) data is presented in Section \ref{sec:gt}.

\section{Extraction of Gamow-Teller strengths and dipole cross sections}\label{sec:gt}

To gain more insight into the various multipole contributions to the spectra extracted from the $^{150}$Nd($^3$He,$t$) and $^{150}$Sm($t$,$^3$He) experiments, Multipole Decomposition Analyses (MDA) \cite{MON87} were performed. In the MDA, measured differential cross sections were fit to a linear combination of theoretical angular distributions associated with different units of angular momentum transfer.

The DWBA code FOLD \cite{FOLD} was used to calculate differential cross sections.  Using this code, form factors were constructed by double-folding the effective nucleon-nucleon interaction at 140 MeV/u from Refs. \cite{LOV81,FRA85} over the transition densities of the projectile-ejectile (i.e. $^{3}$He--$t$ or $t$--$^{3}$He) and target-residue (i.e. $^{150}$Nd--$^{150}$Pm or $^{150}$Sm--$^{150}$Pm) systems. One-body transition densities (OBTDs) for the target-residue system were generated in a normal-mode formalism \cite{HOF95} using the code \textsc{NORMOD} \cite{NORM}. In this formalism, the set of OBTDs generated for a particular operator that connects the initial and final states exhausts the full strength associated with that operator within the model space, i.e. 100\% of the corresponding non-energy weighted sum rule (NEWSR) is exhausted. For $^{150}$Nd, occupation numbers of single-particle states were calculated using the Skyrme SK20 potential \cite{BRO98}: protons filled all orbits up to and including $1g_{7/2}$, and two protons were in the $2d_{5/2}$ orbit. Neutrons filled all orbits up to and including the $2f_{7/2}$ orbit. For $^{150}$Sm, a similar procedure was followed, but of the four protons that were calculated to be in the $2d_{5/2}$ orbit, two were instead placed in the $1h_{11/2}$ orbit; otherwise GT transitions would have been completely Pauli-blocked. All neutron orbits below the $2f_{7/2}$ orbit were filled, and six neutrons were in the $2f_{7/2}$ orbit itself.
Radial wave functions were calculated with Woods-Saxon potentials, for which the well-depths were adjusted such that the single-particle binding energies matched those calculated using the above-mentioned SK20 interaction.
For the $t$ and $^{3}$He particles, radial densities obtained from Variational Monte-Carlo calculations \cite{WIR05} were used and all protons and neutrons were assumed to be in the $1s_{1/2}$ orbit.

\begin{figure}
  \centering
    \includegraphics[scale=1.0]{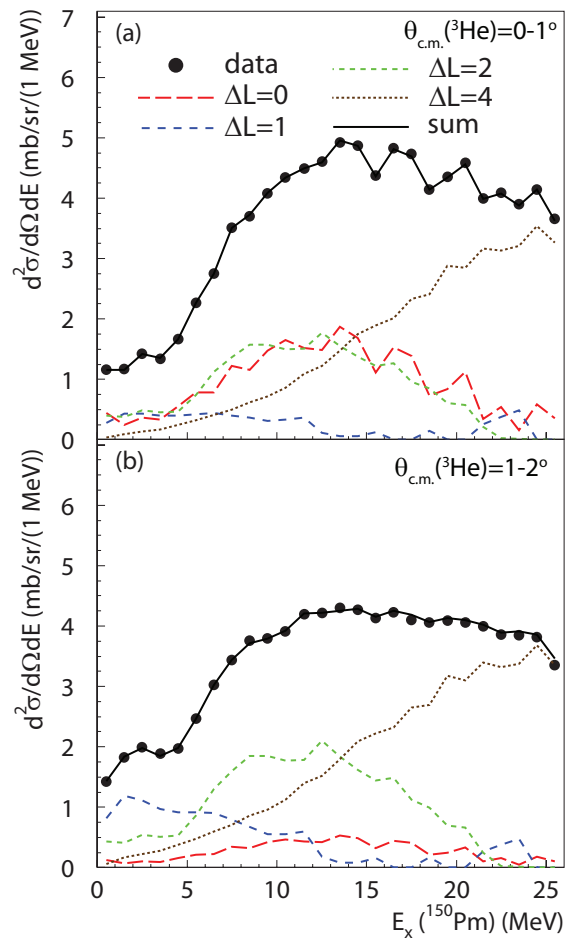}
\caption[]{(color online) MDA results for the 0$^\circ$-1$^\circ$ and 1$^\circ$-2$^\circ$ angular bins in the $^{150}$Sm($t$,$^3$He) experiment. Contributions from $\Delta L=2$ and $\Delta L=4$ transitions are strongly biased due to the absence of a $\Delta L=3$ component in the MDA (see text). In both (a) and (b), the results in each 1-MeV wide bin from the MDA for each multipole are connected by lines, rather then showing the individual points.}
\label{Sm_md_summary_1}
\end{figure}

The calculated form factors served as input for DWBA calculations. The optical potential parameters determined by fitting $^{150}$Nd($^{3}$He,$^{3}$He) elastic scattering data as discussed in Section \ref{sec:ndexp} were also used. Following Ref. \cite{WER89}, the depths of the optical potentials for the tritons in the outgoing channel were scaled from those for the $^{3}$He particles in the incoming channel by a factor of 0.85. For the analysis of the $^{150}$Sm($t$,$^{3}$He) reaction, the optical potentials for the incoming and outgoing channels were interchanged from those for the $^{150}$Nd($^{3}$He,$t$) reaction.

Angular distributions were calculated for transitions with orbital angular momentum transfer $\Delta L$ = 0, 1, 2, 3, and 4 as shown in Fig. \ref{Nd_shapes} for the $^{150}$Nd($^{3}$He,$t$) reaction. These distributions, calculated at the appropriate reaction $Q$-value for a particular transition or excitation energy bin, became the base distributions for the MDA.

\begin{figure}
  \begin{center}
       \includegraphics[width=0.5\textwidth]{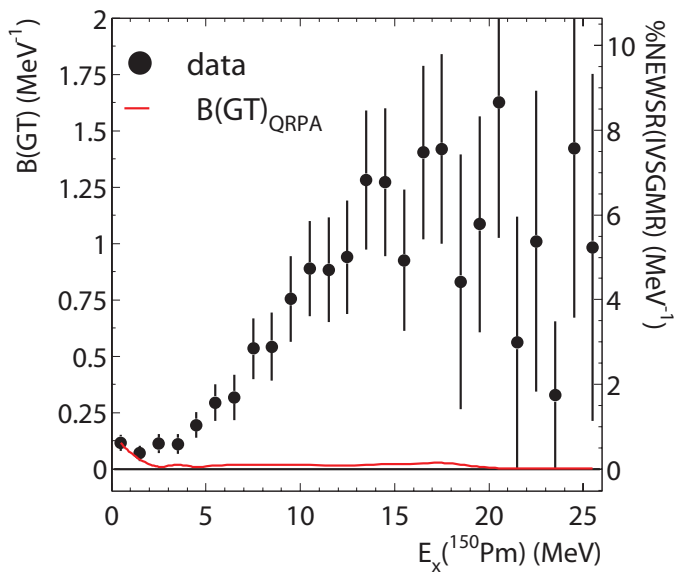}
  \end{center}
  \caption[]{(color online) Strengths associated with $\Delta L=0$ transitions in the $^{150}$Sm($t$,$^3$He) data interpreted as fully due to GT transitions (vertical scale on left-hand side axis) and fully due to the excitation of the IVSGMR (vertical scale on the right-hand axis). The solid red line represents the calculated GT strength distribution in QRPA. Based on the comparison with the QRPA calculations and simple considerations based on the Pauli principle, it is concluded that the vast majority of the observed $\Delta L=0$ strength is due to the excitation of the IVSGMR, except at excitation energies below 2 MeV.}
  \label{Sm_gt_th}
\end{figure}

\subsection{Analysis of the $^{150}$Nd($^3$He,$t$) experiment}\label{anand}
Individual peaks below $E_{x}(^{150}$Pm)=2 MeV in Fig. \ref{Nd_cs1}(b) were analyzed on a case-by-case basis.
Only a few peaks were completely separable from others because of the high density of states. Therefore, the background yield below each non-separable peak was subtracted by fitting the spectra in each angular bin with a linear combination of a Gaussian-shaped peak (or multiple peaks, if not separable) and a ``background'' represented by a function linear in excitation energy. The differential cross sections for the IAS, located at $E_{x}(^{150}$Pm)=14.35 MeV, were extracted in a similar manner.

Angular distributions for each peak were then decomposed into contributions belonging to different units of angular momentum transfer. Three examples are shown in Fig. \ref{Nd_low2}. Fig. \ref{Nd_low2}(a) shows the differential cross section and MDA for the transition to the ground state of $^{150}$Pm. The ground state is dominated by a dipole ($\Delta L$=1) contribution.  A significant $\Delta L$=3 contribution is also observed. Given the uncertainties in the MDA, the $\Delta L$=2 contribution is probably insignificant, although the presence of another state at very low excitation energies cannot be completely excluded. Barrette {\emph{et al.}} \cite{BAR69} studied the decay of $^{150}$Pm to $^{150}$Sm and assigned a very tentative J of 1 to the ground state of $^{150}$Pm. However, unless two levels exist at energies too close to separate in the current work, our results suggest that the ground state of $^{150}$Pm may have a spin-parity (J$^\pi$) of 2$^-$. A $1^{+}$ assignment can be ruled out.

The strongly-populated state at 0.11 MeV (see Fig. \ref{Nd_low2}(b))  has a large $\Delta$L=0 component associated with a GT transition and the population of a $1^{+}$ level. The $\Delta L=2$ contribution to this peak is likely associated with the excitation of the same $1^{+}$ level, although the additional presence of a separate $2^{+}$ or $3^{+}$ state in close proximity to the $1^{+}$ state cannot be excluded. A significant $\Delta L=1$ contribution is also found, indicating the presence of a $0^{-}$, $1^{-}$ or $2^{-}$ level.  The differential cross section at $0^{\circ}$ for the $\Delta$L=0 contribution to this peak was extracted from the fit. Its value is 0.565$\pm$0.085 mb/sr, where the error includes statistical and systematic uncertainties in the MDA.

Fig. \ref{Nd_low2}(c) shows the differential cross section for the IAS. As expected, its angular distribution is consistent with a pure $\Delta L=0$ distribution. The extracted differential cross section at $0^{\circ}$ was $9.9\pm0.3$ mb/sr, where the error represents the statistical and fitting errors only. On the basis of DWBA calculations (see e.g. Ref. \cite{ZEG06} Eq. (2)), the differential cross section was extrapolated to vanishing linear momentum transfer ($q=0$) with a value of $9.6\pm0.3$ mb/sr. Since the Fermi transition strength associated with the IAS exhausts the Fermi sum rule (N-Z)=30, the Fermi unit cross section was found to be $\hat{\sigma}_{F}=0.32\pm0.01$ mb/sr. This is about 10\% different from the value calculated using Eq. (\ref{eq:uf}), typical for deviations seen between the phenomenological equation and extracted Fermi unit cross sections for other target masses \cite{ZEG07}.

Differential cross sections associated with the $\Delta$L=0 component were likewise extracted for all distinguishable peaks (21) below an excitation energy of 2.1 MeV. The differential cross sections at $0^{\circ}$ were then extrapolated to $q=0$ and Eq. (\ref{eq:ugt}) used to determine the associated GT strengths. These strengths are listed in Table \ref{tb:Nd_bgt} and shown in Fig. \ref{dis_Bgt}. In addition to errors associated with statistics, the subtraction of background and fitting errors, a 15\% estimated error due to systematic uncertainties in the MDA analysis was included. These errors are related to the assumptions made in the DWBA calculations and the fact that only a limited number of angular momentum transfers could be considered given the angular range over which data were acquired. These systematic errors were assumed to be uncorrelated for each of the 21 transitions listed in Table \ref{tb:Nd_bgt}, providing a summed strength below 2.1 MeV of $0.37\pm0.03$. A correlated error arises from the uncertainty in the unit cross section. It was estimated to be 10\%, based on the above-mentioned difference between the extracted and phenomenological Fermi unit cross section.

Due to the high level density, peak-by-peak analysis was not possible at excitation energies above 2 MeV. Moreover, GT and other multipole strengths can be contained in weakly-excited states below 2 MeV for which peaks cannot be discerned in the spectrum. Therefore, a MDA analysis was performed for 1-MeV wide bins in excitation energy up to 30 MeV. In addition, for the combined evaluation of the $^{150}$Sm($t$,$^{3}$He) and $^{150}$Nd($^{3}$He,$t$) data at low-excitation energies of relevance for the $2\nu\beta\beta$ matrix element (see Section \ref{sec:me}), a MDA was also performed for the three 0.3-MeV wide energy bins between 2.1 and 3 MeV.
The MDA procedure applied to the 0.3 and 1-MeV wide energy bins was otherwise identical to the one applied for the low-lying peaks discussed above. Since the focus of the current analysis is on excitations associated with $\Delta L=0$ and $\Delta L=1$, the results of the MDA are shown as a function of excitation energy in Fig. \ref{Nd_md_summary} for the angular bins $0^{\circ}-0.5^{\circ}$ (Fig. \ref{Nd_md_summary}(a)) and $1.5^{\circ}-2.0^{\circ}$ (Fig. \ref{Nd_md_summary}(b)), where the angular distributions associated with $\Delta L=0$ and $\Delta L=1$ peak, respectively. The contribution from the IAS to the spectra was subtracted prior to the MDA. Therefore, although the excitation of the IAS produces a strong peak in the 14--15 MeV bin in the data, its contribution is not reflected in the results from the MDA.

\begin{figure}
  \centering
    \includegraphics[scale=1]{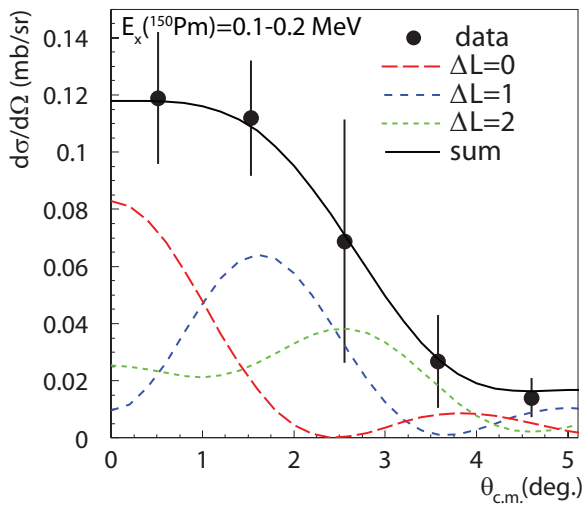}
    \caption[]{(color online) Results of MDA fit for the 0.1-0.2 MeV excitation energy bin in $^{150}$Sm($t$,$^3$He)$^{150}$Pm.  The $\Delta L=0$ angular distribution at 0$^\circ$ has a cross section of 0.08 mb/sr $\pm$ 0.03 mb/sr. The error bars on the data points are due to statistical errors and systematic errors in the subtraction of background (see Section \ref{subsec:smexp}).}
    \label{Sm_md_1_100}
\end{figure}

Aside from the IAS, the $\Delta L=0$ contribution that dominates the spectrum at forward scattering angles (Fig. \ref{Nd_md_summary}(a)) exhibits a characteristic peak due to the excitation of the GTR, which is centered around $\sim 15$ MeV. Long tails extend to lower and higher excitation energies. The GT strength at excitation energies below the main peak have been studied in detail for tin isotopes \cite{JAN93, PHA95}. Whereas the main peak of the GTR was attributed mostly to ``direct spin-flip'' excitations, associated with neutrons from orbits with $j = l+\frac{1}{2}$ being exchanged for protons in the spin-orbit partner orbits with $j = l-\frac{1}{2}$, the strength at lower excitation energies was attributed to ``core polarization spin-flip'' ($j = l \pm \frac{1}{2} \rightarrow j = l \pm \frac{1}{2}$) and ``back spin-flip'' ($j = l - \frac{1}{2} \rightarrow j = l + \frac{1}{2}$) modes. The latter two contributions, sometimes referred to as pygmy resonances \cite{GUB89,ROD03},  become strong in the $\Delta T_{z}=-1$ direction for relatively neutron-rich nuclei, and the $^{150}$Nd($^{3}$He,$t$) reaction is such a case.

A significant amount of $\Delta L=0$ strength is found at excitation energies above the main GTR peak. It closely resembles the results from ($p$,$n$) experiments on medium-heavy nuclei, such as $^{90}$Zr \cite{YAK05} and $^{116}$Cd \cite{SAS09}. This high-lying $\Delta L=0$ strength can be attributed to two sources, which cannot be separated in the present experiment. The first source is related to the well-known ``quenching'' of Gamow-Teller strength: it is found that only about 60\% of the strength associated with the
Ikeda sum rule,
\begin{equation}
S_{\beta^{-}}(GT)-S_{\beta^{+}}(GT)=3(N-Z),
\label{sumrule}
\end{equation}
is found in the GTR and states at lower excitation energies \cite{GAA81,GAA85}. This is partially explained by mixing between $1p$--$1h$ and $2p$--$2h$ configurations via the strong tensor interaction \cite{HYU80,ARI99}, which relocates strength to excitation energies beyond the GTR. This mechanism is experimentally confirmed by results from $^{90}$Zr($p$,$n$) and  $^{90}$Zr($n$,$p$) experiments \cite{YAK05}. Although in the present experiment only excitation energies up to 30 MeV are covered, the relocated GT strength is likely responsible for the majority of $\Delta L=0$ contributions seen in  Fig. \ref{Nd_md_summary}(a) at excitation energies above 20 MeV. The second source of high-lying $\Delta L=0$ strength is the excitation of the IVSGMR, discussed in Section \ref{sec:intro}. Although it peaks at an excitation energy above 30 MeV in the $\Delta T_{z}=-1$ direction, it is expected to contribute to the spectrum below that energy due to its large width. Based on results from the $^{208}$Pb($^{3}$He,$t$) reaction \cite{ZEG03}, its contribution to the strength associated with $\Delta L=0$ below 30 MeV is expected to be small, but could be partially responsible for the small increase of the monopole strength observed above 25 MeV. However, without data at higher excitation energies, this could not be investigated in further detail.  In addition, small systematic uncertainties in the MDA, in combination with an increasing factor for the extrapolation of the $\Delta L=0$ cross section at $0^{\circ}$ to $q=0$, could contribute to the artificial increase of strength at higher excitation energies.

Differential cross sections at $0^{\circ}$ associated with the $\Delta L=0$ contributions to the excitation energy spectrum were extrapolated to $q=0$ and Eq. (\ref{eq:ugt}) was applied to estimate the corresponding GT strength. These strengths, for each 1-MeV wide bin in excitation energy, are provided in Table \ref{tb:Nd_bgt} and also shown in Fig. \ref{Nd_gt_th}. The GT strengths for the three 0.3-MeV wide bins between 2.1 and 3.0 MeV (which coincide with the onset of the above-mentioned Pygmy resonances) are given in the right-hand column of that same table and displayed in Fig. \ref{dis_Bgt}.
The total GT strength observed at excitation energies below 30 MeV is $50.0\pm1.7\pm5.0$, where the first error is due to uncorrelated statistical and systematic uncertainties in each 1-MeV wide bin and the second error is due to the uncertainty in the GT unit cross section that affects all extracted strengths equally. The sum rule of Eq. (\ref{sumrule}) gives a value of 90, assuming $S_{\beta^{+}}=0$, so that the total extracted GT strength corresponds to an exhaustion of $55\pm2\pm6$\%. The summed GT strength in the first two 1-MeV wide bins is $0.46\pm0.05$, whereas the peak-by-peak analysis gives $0.37\pm0.03$. Ignoring correlated errors because they are the same for both methods, the slightly higher value in the former method indicates the presence of some GT strength not clearly associated with peaks in the spectrum below 2 MeV.

As shown in Fig. \ref{Nd_md_summary}(b), the spin-dipole resonance (IVSGDR) is evident at angles of 1.5$^\circ$-2$^\circ$ and peaks at an excitation energy of 22 MeV. Significant dipole contributions to the excitation energy spectrum  are also found at lower excitation energies. As mentioned, a proven proportionality between strengths and differential cross sections for dipole transitions is lacking. However, the shape of the extracted $\Delta L=1$ distribution was compared with the theoretical strength distribution calculated in QRPA, as discussed in Section \ref{sec:comp}.

Given the limited angular range covered in the present experiment, contributions to the spectra from transitions associated with $\Delta L\geq 2$ were not investigated in detail. The larger the transfer of angular momentum, the larger the uncertainties in the associated contributions extracted from the data, due to the absence of assumed contributions from transitions with $\Delta L \geq 5$ in the MDA. Although the extraction of $\Delta L=2$ contributions should be relatively reliable compared to those associated with $\Delta L=3$ and 4 , the lack of any specific features in the spectrum associated with this transition makes it hard to gain insight in the quality of the extracted distribution, even on the qualitative level.

\subsection{Analysis of the $^{150}$Sm($t$,$^3$He) experiment}\label{anasm}
The analysis of the $^{150}$Sm($t$,$^3$He) data was similar to that of the $^{150}$Nd($^3$He,$t$) data, but complicated by lower statistics and poorer excitation-energy resolution. Because of the lower statistics, the data set could only be subdivided into 5 separate $1^{\circ}$-wide scattering-angle bins, limiting the MDA to at most 4 different angular momentum components. An analysis with contributions associated with $\Delta L=0,1,2$ and 3 resulted in poor fitting results at the largest angles, indicating the necessity of including a contribution due to $\Delta L=4$ transitions, which could only be accomplished by excluding the $\Delta L=3$ contribution in the fits. The use of a $\Delta L=4$ component instead of $\Delta L=3$ component improved the overall quality of the fits, but strongly affected the extracted strength distribution for transitions associated with $\Delta L=2$. However, the results from the MDA for the $\Delta L=0$ and 1 contributions to the spectrum were not strongly affected (compared to the statistical uncertainties) by the choice of which higher multipole was included. The results presented in this work are from the MDA with $\Delta L=0,1,2$ and 4 contributions, but we stress that by leaving out the $\Delta L=3$ contribution in the MDA, the results for $\Delta L=2$ and 4 contributions are heavily biased.

Fig. \ref{Sm_md_summary_1} shows the MDA results as performed for 1-MeV wide bins for the excitation-energy spectrum up to 26 MeV and scattering angles between $0^{\circ}$ and $1^{\circ}$ (Fig. \ref{Sm_md_summary_1}(a)) and $1^{\circ}$ and $2^{\circ}$ (Fig. \ref{Sm_md_summary_1}(b)). Transitions associated with $\Delta L=0$ peak at $0^{\circ}$ and thus appear strongest in Fig. \ref{Sm_md_summary_1}(a). A broad resonance-like structure is observed, centered around an excitation energy of about 13 MeV. Dipole transitions, which peak at $\sim 1.5^{\circ}$, are seen predominantly at low excitation energies in Fig. \ref{Sm_md_summary_1}(b) and are nearly absent above 15 MeV. The steady decrease of the $\Delta L=2$ contributions above 15 MeV and steady increase of the $\Delta L=4$ contributions are partially caused by the absence of a $\Delta L=3$ component in the fit and likely artificial, as mentioned above.

As discussed in Section \ref{sec:intro}, GT transitions from $^{150}$Sm to $^{150}$Pm are expected to be strongly Pauli blocked. On the other hand, the broad IVSGMR is expected to peak at 15-20 MeV, similar  to the $\Delta L=0$ distribution extracted from the data. To gain more insight into the nature of the observed $\Delta L=0$ strength, we tested two hypotheses: one assumed that all $\Delta L=0$ strength was due to Gamow-Teller transitions and the second that assumed it was entirely due to the excitation of the IVSGMR. To test the first hypothesis, the measured differential cross sections for the $\Delta L=0$ contribution in each 1-MeV wide bin were extrapolated to $q=0$, and the GT strengths were extracted by using Eq. (\ref{eq:ugt}). This method is identical to that applied for the extraction of GT strength from the $\Delta L=0$ distribution in the $^{150}$Nd($^{3}$He,$t$) reaction. The results are shown in Fig. \ref{Sm_gt_th}, in which the vertical scale on the left-hand axis refers to the GT strength extracted on the basis of the above hypothesis. The error bars in this figure include statistical errors, errors associated with the subtraction of background and a 15\% estimated error due to systematic uncertainties in the MDA analysis.
The relatively large error margins for the data points at high excitation energies are due to the small magnitudes of $\Delta L=0$ strength extracted in the MDA in combination with the relatively large multiplicative factor associated with the extrapolation to $q=0$ at large excitation energies: even minute cross sections extracted from the MDA for $\Delta L=0$ transitions represent a relatively large amount of strength. The summed GT strength up to 26 MeV equals $20\pm 2\pm 2$, where the first error includes statistical and systematic errors in the MDA (which were assumed to be uncorrelated between 1-MeV wide energy bins) and the second error refers to the systematic uncertainty in the unit cross section. This large value exceeds by far the amount of GT strengths observed in other ($n$,$p$)-type experiments \cite{YAK05,HEL97,RAY97,RAK05,SAS09} on medium-heavy nuclei. Moreover, Pauli-blocking of GT transitions is expected to be stronger for $^{150}$Sm than for the nuclei studied in those experiment. Therefore, it is not plausible that a large fraction of the extracted $\Delta L=0$ strength is associated with Gamow-Teller transitions.

For the second hypothesis, the differential cross sections extrapolated to $q=0$ were used to calculate the percentage by which the NEWSR for the IVSGMR was exhausted. It was assumed that the percentage of exhaustion of the NEWSR for the IVSGMR is proportional to the cross section at $q=0$: i.e. that a unit cross section $\hat{\sigma}_{IVSGMR}$ exists that serves the same purpose as $\hat{\sigma}_{GT}$ ($\hat{\sigma}_{F}$) for the GT (Fermi) $\Delta L=0$ transitions. The value of $\hat{\sigma}_{IVSGMR}$ was determined by calculating the ratio of  $\hat{\sigma}_{IVSGMR}$ to $\hat{\sigma}_{GT}$ in DWBA and rescaling $\hat{\sigma}_{IVSGMR}$ by the same factor needed to match the calculated value of $\hat{\sigma}_{GT}$ in DWBA to the empirical value from Eq. (\ref{eq:ugt}). We found that $\hat{\sigma}_{IVSGMR}=0.45$ mb/sr per 1\% of the full NEWSR strength of the IVSGMR (100\% of the NEWSR corresponds to a strength of 1433 fm$^{4}$ as calculated in the normal-modes formalism). The extracted exhaustion of the NEWSR for the IVSGMR is also shown in Fig. \ref{Sm_gt_th}: the relevant scale is defined on the right-hand side. The summed exhaustion is $106\pm 11 \pm 11$\%, where the error bars have meanings similar to those for the GT strength above. This number is inflated by as much as 20\% due to a small contribution from the IVGMR (estimated at $\lesssim 5$\%), the presence of some GT strength and the possible misinterpretation of small and perhaps spurious $\Delta L=0$ contributions at high excitation energies which add significantly to the strength observed due to the extrapolation to $q=0$.
Nevertheless, this result provides strong evidence for the excitation of the IVSGMR. The large error bars at high excitation energies make it difficult to extract accurate resonance parameters, but the approximate peak location of 15 MeV and width of 10 MeV are consistent with the expectation for the IVSGMR.

The large contributions from the IVSGMR to the spectrum make it hard to extract GT strength of interest for $\beta\beta$-studies. Nevertheless, the spectrum below 3 MeV was studied in more detail to search for isolated transitions that could be associated with GT transitions. The contributions from the IVSGMR are expected to be small at these low excitation energies and not expected to exhibit isolated peaks. The excitation-energy region below 3 MeV was divided into bins of 300-keV and a MDA performed for each bin. The extracted $\Delta L=0$ contributions were assumed to be due to Gamow-Teller transitions and GT strengths were deduced following the procedure described above.
The results are shown in Fig. \ref{dis_Bgt} (red square markers) and also provided in Table \ref{tb:Sm_bgt}.

To test the SSD hypothesis for $2\nu\beta\beta$ decay, special scrutiny was given to the excitation-energy region below 300 keV; in particular whether a GT transition could be identified to match the excitation of the $1^{+}$ state at 0.11 MeV observed in the $^{150}$Nd($^3$He,$t$) experiment. The GT strength observed in the first 300-keV wide bin was $0.021\pm0.013$. However, upon closer inspection, it was found that that strength was concentrated between 100 and 200 keV and a MDA of that 100-keV wide bin (see Fig. \ref{Sm_md_1_100}) resulted in a $B(GT)$ of $0.021\pm0.008$, whereas in the neighboring 100-keV wide bins the GT strength was consistent with zero. Taking into account the systematic error of 50 keV in the determination of the excitation energy in the $^{150}$Sm($t$,$^{3}$He) experiment, we concluded that a $1^{+}$ state (or more than a single state) is excited with a B(GT) of $0.021\pm0.008$ that is located at an excitation energy between 50 and 250 keV, as indicated in Table \ref{tb:Sm_bgt} and by the horizontal error bars in Fig. \ref{dis_Bgt}. Further concentrations of GT strength were observed between 0.6 and 1.2 MeV and between 2.1 and 2.4 MeV. GT strengths in other 300 keV-wide bins below 3 MeV were consistent with zero.

\begin{table}
\caption[]{\label{tb:Sm_bgt}GT transition strengths extracted from the $^{150}$Sm($t$,$^{3}$He) experiment at excitation energies in $^{150}$Pm below 3 MeV. Strengths are extracted in 300-keV wide bins, but for the lowest bin it was possible to locate the strength more accurately, as indicated in the Table and discussed in detail in the text.}
\begin{ruledtabular}
\begin{center}
\begin{tabular}{cc}
E$_x$($^{150}$Pm) (MeV)& $B(GT)$ \\ \hline
\noalign{\smallskip}
    0.05--0.25 & 0.021 $\pm$ 0.008 \\
    0.3--0.6 & 0.017 $\pm$ 0.017 \\
    0.6--0.9 & 0.057 $\pm$ 0.021 \\
    0.9--1.2 & 0.048 $\pm$ 0.021 \\
    1.2--1.5 & 0.006$^{+0.017}_{-0.006}$ \\
    1.5--1.8 & 0.019 $\pm$ 0.019 \\
    1.8--2.1 & 0.019 $\pm$ 0.019 \\
    2.1--2.4 & 0.089 $\pm$ 0.030 \\
    \noalign{\smallskip}
    2.4--2.7 & 0.013$^{+0.022}_{-0.013}$ \\
    \noalign{\smallskip}
    2.7--3.0 & 0.007$^{+0.022}_{-0.007}$ \\
\end{tabular}
\end{center}
\end{ruledtabular}
\end{table}

\section{Comparison with QRPA calculations}\label{sec:comp}
Motivated by plans to study $\beta\beta$-decay of $^{150}$Nd at the SNO+ experiment \cite{KRA10} and other facilities, a significant effort has gone into improving theoretical calculations relating to the $0\nu\beta\beta$ decay of $^{150}$Nd. Some of the most recent works apply the proton-neutron QRPA and takes into account the relevant nuclear deformations \cite{YOU09,FAN10,FAN10a,FAN11}.
We compared the theoretical results for GT and spin-dipole distributions based on that same formalism with our experimental results. For details concerning the calculations, we refer to the above-mentioned references (in particular Refs. \cite{FAN10a,FAN11}, which hold the latest results) and restrict ourselves to giving the key parameters in the QRPA. The geometric deformation parameters [$\beta_{2}=0.240(0.153)$ for $^{150}$Nd ($^{150}$Sm)] for the deformed Woods-Saxon mean-fields were adjusted so that the empirical deformation parameters [$\beta=0.29(0.19)$ for $^{150}$Nd ($^{150}$Sm)] deduced from experimental $B(E2)$ values were reproduced in the calculations. The QRPA particle-hole  renormalization factor of the residual interaction $g_{ph}=0.9$ was fixed by fitting the experimental  position of the GTR in $^{76}$Ge. The particle-particle renormalization factor $g_{pp}$ was set to 1.0, so that the experimental value for the $2\nu\beta\beta$ half-life is reproduced. In this adjustment a quenching factor $g_{eff}= 0.75$ was taken into account, and the same quenching factor was also applied to the calculated GT strength distributions (thus scaled by $(0.75)^{2}$ before comparing with the experimental results).
Since spreading effects are not included in the QRPA calculations, they produce a large number of isolated states. To compare these with our data, the calculated strengths were convoluted with Gaussians ($\sigma$=2 MeV) for strengths located above the threshold for decay by particle emission. For excitation energies below the threshold for decay by particle emission, strengths were convoluted with Gaussian distributions having a width equal to the experimental energy resolution and summed over 1-MeV wide-bins.

Fig. \ref{Nd_gt_th} shows the comparison between the experimental GT strength distribution from the $^{150}$Nd($^{3}$He,$t$) experiment and the associated QRPA calculation. The location of the GTR is well reproduced by the theory, as well as the presence of a tail towards lower excitation energies due to the above-mentioned pygmy resonances. However, the GT strength extracted at excitation energies above the GTR in the data is not reproduced by the QRPA results, as mixing between $1p$--$1h$ and $2p$--$2h$ configurations and other possible effects that quench the GT strength at lower excitation energies were not included in the calculations. Therefore, even though the theoretical GT strength distribution was scaled by a quenching factor, the strength removed at and below the main GTR is not recovered at higher excitation energies. We also note that the QRPA calculations predict significantly more strength at very low excitation energies (below 2 MeV) than seen in the data.

From the comparison between the QRPA calculations for the GT strength distribution in the $\Delta T_{z}=+1$ direction and the results from the $^{150}$Sm($t$,$^{3}$He) experiment shown in Fig. \ref{Sm_gt_th}, the strong effects of Pauli blocking become very clear. If the experimentally extracted $\Delta L=0$ contributions from the data at excitation energies above 2 MeV were interpreted as GT excitations (left-hand vertical scale of Fig. \ref{Sm_gt_th}), the total strength would exceed the theoretical prediction by more than a factor of 20. Therefore, this comparison strongly supports the interpretation of the extracted $\Delta L=0$ yield as due to the excitation of the IVSGMR.

Spin-dipole transitions involving intermediate transitions to $0^{-}$, $1^{-}$ and $2^{-}$ states are predicted to contribute strongly to the nuclear matrix element for the $0\nu\beta\beta$ decay of $^{150}$Nd \cite{FAN11}. Therefore, besides the GT strength distributions, the extracted dipole distributions were also compared with the QRPA calculations.  The comparison is qualitative only because a proportionality between strength and cross section has not been established for dipole excitations. Nevertheless, the comparison between theoretical strengths and experimental cross sections provides some insight into the quality of the QRPA calculations and is thus included in the present work.

\begin{figure}
  \begin{center}
       \includegraphics[scale=1.0]{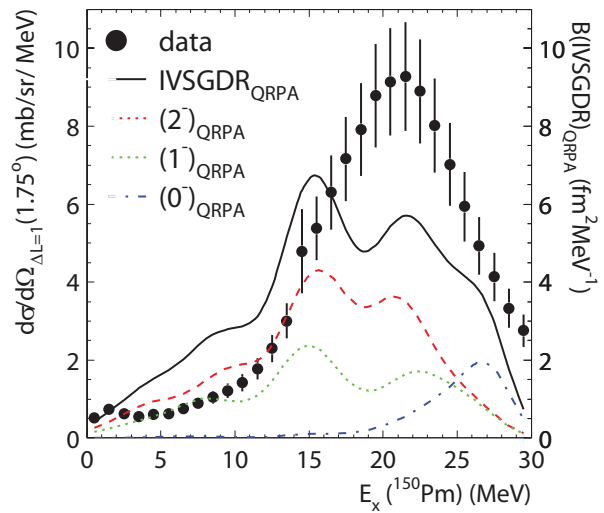}
  \end{center}
  \caption[]{(color online) Differential cross sections associated with dipole transitions extracted from the $^{150}$Nd($^{3}$He,$t$) data (units on left axis) are compared with the strength distribution predicted in QRPA (units right axis). The separate $0^{-}$, $1^{-}$ and $2^{-}$ components that contribute to the total dipole strength predicted in QRPA are shown as well. The scales on the vertical axes are adjusted in order to facilitate an easy comparison between the data and the theoretical calculations, but are otherwise unrelated. The error bars on the data points reflect statistical uncertainties, as well as errors in the MDA.}
  \label{Nd_dip_th}
\end{figure}

Fig. \ref{Nd_dip_th} compares the spin-dipole strength distribution calculated in QRPA and the experimentally extracted differential cross sections at scattering angles between $1.5^{\circ}$ and $2.0^{\circ}$ associated with $\Delta L=1$ excitations from the $^{150}$Nd($^{3}$He,$t$) experiment. The latter include a minor ($\lesssim 5$\%) contribution from non-spin-transfer isovector dipole transitions, which were not included in the theoretical calculation. Besides the total QRPA spin-dipole strength, the separate contributions from transitions to $0^{-}$, $1^{-}$ and $2^{-}$ states are included in Fig. \ref{Nd_dip_th} as well. The theoretical distribution exhibits more structure and is broader than the dipole strength extracted from the data.

\begin{figure}
  \begin{center}
       \includegraphics[width=0.5\textwidth]{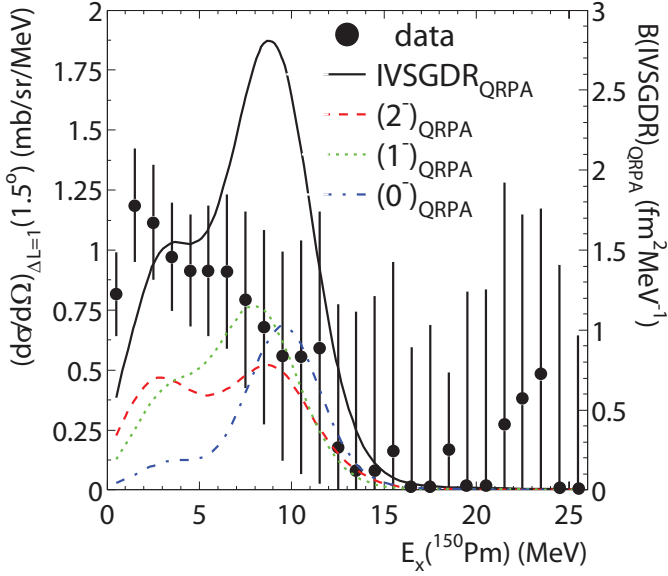}
  \end{center}
  \caption[]{(color online) Differential cross sections associated with dipole transitions extracted from the $^{150}$Sm($t$,$^{3}$He) data (units on left axis) are compared with the strength distribution predicted in QRPA (units on right axis). The separate $0^{-}$, $1^{-}$ and $2^{-}$ components that contribute to the total dipole strength predicted in QRPA are shown as well. The scales on the vertical axes are adjusted in order to facilitate an easy comparison between the data and the theoretical calculations, but are otherwise unrelated. The error bars on the data points reflect statistical uncertainties, as well as errors in the MDA.}
  \label{Sm_dip_th}
\end{figure}

In Fig. \ref{Sm_dip_th}, the dipole distribution extracted from the $^{150}$Sm($t$,$^{3}$He) data is compared to the spin-dipole strength distribution from the QRPA calculations. Both theory and experiment place the dipole strength at excitation energies below 15 MeV, although the QRPA calculations peak at $\sim 9$ MeV, whereas the experimental distribution peaks at lower excitation energies.

\section{Calculation of the 2$\nu\beta\beta$ decay matrix element assuming SSD}\label{sec:me}
Calculation of the nuclear matrix element for either 2$\nu$- or 0$\nu\beta\beta$ decay of $^{150}$Nd relies on more than the multipole strength distributions in the intermediate nucleus $^{150}$Pm, because phase factors for adding contributions from individual transitions in the two legs of $\beta\beta$ decay can be different. However, if the SSD hypothesis for 2$\nu\beta\beta$ is valid, only one transition through a single intermediate state matters, and the presence of a strong 1$^+$ state at 0.11 MeV in the $^{150}$Nd($^3$He,$t$) data makes 2$\nu\beta\beta$ decay of $^{150}$Nd a good test case.

The main complication for testing the SSD hypothesis from the current data is the ambiguity about the nature of the GT strength observed in the $^{150}$Sm($t$,$^{3}$He) in the first 300-keV wide energy bin. Although the detailed analysis described above made it possible to restrict the location of the GT strength to the region between 50 and 250 keV, it is not guaranteed that the strength (solely) corresponds to the excitation of the state seen at 0.11 MeV, or (partially) corresponds to the much weaker state observed at 0.19 MeV in the $^{150}$Nd($^3$He,$t$) data set (see Fig. \ref{dis_Bgt}). Therefore, we can only provide an upper limit for the nuclear matrix element for $2\nu\beta\beta$ decay (and, consequently, a lower limit for the half-life) under the assumption that SSD holds and all GT strength observed between 50 and 250 keV in the $^{150}$Sm($t$,$^{3}$He) data corresponds to the population of the $1^{+}$ state seen at 0.11 MeV in the $^{150}$Nd($^3$He,$t$) experiment.
Eqs. (\ref{eq:dbd_2nu}) and (\ref{eq:dbd_GTME}) were used to calculate $M_{GT}^{2\nu}$ and 2$\nu\beta\beta$ decay half-life from the extracted GT strengths. The phase-space factor $G^{2\nu}=3.1\times 10^{-17}$ yr$^{-1}$MeV$^{2}$ was taken from Ref. \cite{SUH98}.

\begin{figure}
  \centering
    \includegraphics[scale=0.85]{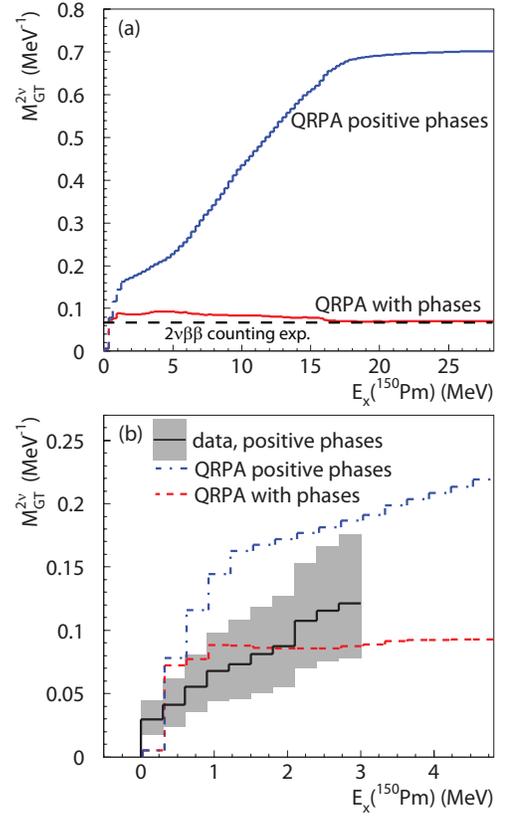}
    \caption[]{(color online)(a) Running sums of the $2\nu\beta\beta$ matrix element as a function of excitation energy in the intermediate nucleus $^{150}$Pm, as calculated in QRPA. For the blue curve it was assumed that all contributions add coherently (all positive phases), whereas for the red curve phases were taken into account. The dashed line indicates the matrix element calculated based on the measured $2\nu\beta\beta$ half-live of $^{150}$Nd. (b) \emph{Idem}, but zoomed in at low-excitation energies. Also shown is the running sum of the $2\nu\beta\beta$ matrix element based on the experimentally extracted GT strengths from the $^{150}$Sm($t$,$^{3}$He) and $^{150}$Nd($^3$He,$t$) experiments. All phases were assumed to be positive and the shaded area indicates the error margins. Note that this curve reflects an upper limit, as it is assumed that GT strength from the two charge-exchange legs in each 300-keV wide bin belong to the excitation of the same intermediate $1^{+}$ states in $^{150}$Pm.}
    \label{mgt}
\end{figure}

The results of the calculations and the comparison with measured 2$\nu\beta\beta$ decay half-lives are provided in Table \ref{tb:dis_me}. The calculated half-life under the above-mentioned conditions (second column of Table \ref{tb:dis_me}) is larger by more than a factor of 4 than the recommended values in Refs. \cite{BAR10} (column 4) and \cite{PRI10} (column 5) based on direct counting experiments. Given that this estimate provides a lower-limit for the half-life, it can be concluded that SSD can be excluded for the 2$\nu\beta\beta$ decay of $^{150}$Nd at the 2$\sigma$ level. We also calculated the half-life based on the assumptions that matrix elements can be calculated for each 300-keV wide excitation-energy bin in $^{150}$Pm and summed coherently. GT strengths extracted from the $^{150}$Sm($t$,$^{3}$He) and $^{150}$Nd($^3$He,$t$) data are, moreover, assumed to populate the same states within these 300-keV wide energy bins.
The result is provided in column 3 of Table \ref{tb:dis_me}. This is also a lower limit for the half-life, but the value is approximately a factor of 4 lower than the values based on the direct counting experiments. Therefore, it can be concluded that the 2$\nu\beta\beta$ decay half-life could be due to a combination of transitions of intermediate $1^{+}$ states at low excitation energies in $^{150}$Pm. We stress, however, that it cannot be excluded that even higher lying intermediate transitions also play a role.

\begin{table*}
  \caption{\label{tb:dis_me}Comparison of extracted 2$\nu\beta\beta$ decay half-lives for $^{150}$Nd based on extracted GT transition strengths from the $^{150}$Sm($t$,$^{3}$He) and $^{150}$Nd($^3$He,$t$) experiments, and evaluated values \cite{BAR10,PRI10} from direct counting experiments. The values calculated from the charge-exchange data are based on the assumption of SSD through a $1^{+}$ state at $E_{x}(^{150}$Pm)=0.11 MeV (second column) or the most coherent superposition of matrix elements below $E_{x}(^{150}$Pm)=3 MeV (third column). 2$\nu\beta\beta$ double Gamow-Teller matrix elements are also provided in the table.}
  \begin{ruledtabular}
  \begin{center}
    \begin{tabular}{lcccc}
          & \multicolumn{2}{c}{Current work} & \multicolumn{2}{c}{$2\nu\beta\beta$ counting experiments} \\
          \cline{2-3} \cline{4-5}
 \noalign{\smallskip}
          & SSD   & $E_{x}$($^{150}$Pm$)<3.0$ MeV\footnotemark[2] & Ref. \cite{BAR10} & Ref. \cite{PRI10} \\
    \cline{2-2} \cline{3-3} \cline{4-4} \cline{5-5}
    $B(GT)$ $^{150}$Sm$\rightarrow$$^{150}$Pm & 0.021$\pm$0.008\footnotemark[1] & See Table \ref{tb:Sm_bgt} & -     & - \\
    $B(GT)$ $^{150}$Nd$\rightarrow$$^{150}$Pm & 0.13$\pm$0.02 & See Table \ref{tb:Nd_bgt} & -     & - \\
    $M^{2\nu}_{GT}$ (MeV$^{-1}$) & 0.028$\pm$0.006 & 0.13$\pm$0.02 & 0.062$\pm$0.003\footnotemark[3] & 0.064$\pm$0.003\footnotemark[3] \\
    $T^{2\nu}_{1/2}$ (yr) & (4.0$\pm$1.7)$\times 10^{19}$ & (2.0$\pm$0.5)$\times 10^{18}$ & (8.2$\pm$0.9)$\times 10^{18}$ & (7.9$\pm$0.7)$\times 10^{18}$ \\
    \end{tabular}
    \end{center}
    \end{ruledtabular}
    \footnotetext[1]{Assumes that all GT strength extracted between 50 keV and 250 keV from the $^{150}$Sm($t$,$^{3}$He) experiment is associated with the excitation of the 0.11 MeV state from the $^{150}$Nd($^{3}$He,$t$) experiment (see text).}
    \footnotetext[2]{Calculated by connecting extracted GT strengths from the $^{150}$Sm($t$,$^{3}$He) and $^{150}$Nd($^{3}$He,$t$) experiments per 300-keV wide excitation-energy bin in $^{150}$Pm and assuming that matrix elements from all bins with $E_{x}$($^{150}$Pm$)<3.0$ MeV  add coherently (see text).}
    \footnotetext[3]{Calculated from the quoted evaluated half-lives and by applying Eq. (\ref{eq:dbd_0nu}).}
\end{table*}

The upper limits for the 2$\nu\beta\beta$ matrix elements extracted from the charge-exchange experiments can be compared with those predicted in the framework of QRPA. Since the matrix elements associated with transitions through different intermediate $1^{+}$ states interfere because of their different phases, it is interesting to visualize the evolution of the summed matrix element as a function of excitation energy in the intermediate nucleus $^{150}$Pm. This is done for the theoretical calculations in Fig. \ref{mgt}(a). The blue curve indicates the running sum in case all phases were set positive, so that all contributions add coherently. The running sum increases steadily up to excitation energies of $\sim 18$ MeV, above which the QRPA GT transition strengths from $^{150}$Sm are near zero (see Fig. \ref{Sm_gt_th}). When the phases are taken into account (red line), the matrix element rises to a level just above that expected based on the experimentally extracted value from the 2$\nu\beta\beta$ counting experiments (dashed black line) and then drops very slowly to the experimental value. Note that the full matrix element calculated in QRPA should match the value extracted from the half-life measurements, since the particle-particle renormalization factor $g_{pp}$ was adjusted to reproduce that value.

Fig. \ref{mgt}(b) shows the same running sums of the matrix element calculated in QRPA, but only up to $E_{x}(^{150}$Pm)=5 MeV. The matrix element based on the 2$\nu\beta\beta$ counting experiment is also indicated (with error bar). The solid black line indicates the running sum of the 2$\nu\beta\beta$ matrix element based on the charge-exchange data, assuming coherent superposition of matrix elements calculated per excitation-energy bin of 300 keV (i.e. the value at 3 MeV equals the upper-limit of $0.13\pm0.02$ listed in Table \ref{tb:dis_me}). Note that this curve provides an upper limit for $M^{2\nu}_{GT}$ (and the error bars are the error in that upper limit) as it was assumed that within each 300-keV wide bin, GT transitions from $^{150}$Nd and $^{150}$Sm populate matching intermediate $1^{+}$ states in $^{150}$Pm. Except for the first 300-keV wide bin, the coherently summed matrix elements extracted from the data fall below those calculated in QRPA, indicating excess GT strength at low-excitation energies in the QRPA calculations. Based on the direct comparison between measured and calculated strength in Section \ref{sec:comp} the excess strength seen at low excitation energies in the QRPA calculations for the GT transitions from $^{150}$Nd compared to the data is the likely cause.
As concluded above, the fact that the value of $M^{2\nu}_{GT}$ obtained from the charge-exchange experiments in the first excitation energy bin falls below the value deduced from the experimental 2$\nu\beta\beta$ decay half-life indicates that the conditions for SSD are not met; transitions via intermediate $1^{+}$ states in $^{150}$Pm at least up to an excitation energy of 1 MeV, and potentially higher, are required to explain the measured 2$\nu\beta\beta$ decay half-life.

\section{Conclusions and Outlook}\label{sec:conc}
We have used the $^{150}$Nd($^3$He,$t$) reaction at $140$ MeV/u and $^{150}$Sm($t$,$^3$He) reaction at $115$ MeV/u to study $\Delta L=0$ and $\Delta L=1$ transitions to $^{150}$Pm, which is the intermediate nucleus for the $\beta\beta$ decay of $^{150}$Nd to $^{150}$Sm. In the $^{150}$Nd($^3$He,$t$) experiment, the GT strengths were extracted for 21 transitions to $1^{+}$ excited states below $E_{x}=2$ MeV. In addition, some GT strength was uncovered in the yield below individual peaks through a multipole decomposition analysis. In particular, the first $1^{+}$ state at 0.11 MeV was strongly excited ($B(GT)=0.13\pm0.02$). At higher excitation energies, the GTR dominates the forward-angle yield, with relatively strong tails to lower and higher excitation energies. The excitation of the IVSGDR is also clearly observed, dominating the response at larger scattering angles.

In the $^{150}$Sm($t$,$^3$He) experiment, the spin-flip monopole contributions to the $^{150}$Pm excitation-energy spectrum far exceeded the level that could be expected from GT excitations. Based on the distribution and magnitude of the $\Delta S=1$, $\Delta L=0$ yield, it was interpreted as the excitation of the $2\hbar\omega$ IVSGMR, for which the empirical evidence has been scant in the $\Delta T_{z}=+1$ direction. With an approximate peak-excitation energy of about 15 MeV, a width of about 10 MeV and near-full exhaustion of the associated NEWSR for this resonance, the results present clear evidence for the excitation of the IVSGMR in the $\Delta T_{z}=+1$ direction.

Although the strong excitation of the IVSGMR makes extraction of GT strength from the $^{150}$Sm($t$,$^3$He) experiment difficult, small low-lying amounts of monopole strength are likely due to GT transitions. By comparing the associated strengths to those extracted in the $^{150}$Nd($^3$He,$t$) experiment, we found that the Single-State Dominance hypothesis for the description of $2\nu\beta\beta$ decay through excitations in the nucleus intermediate to the $\beta \beta$-decay mother and daughter is excluded at the $2\sigma$ level and that higher-lying $1^{+}$ states likely play an important role in describing the $2\nu\beta\beta$ decay half-life. The error in the extracted half-life based on GT transitions from the charge-exchange experiments could be further reduced if higher resolution $\Delta T_{z}=+1$ data were available. Given the weakness of the transitions and the high level density in $^{150}$Pm, however, such a measurement might have to involve the high-resolution detection of $\gamma$--rays to uniquely separate the relevant transitions to $1^{+}$ states from neighboring states.

Recent QRPA calculations, performed for the purpose of calculating matrix elements for $0\nu\beta\beta$ decay of $^{150}$Nd, were tested on their ability to accurately reproduce GT and spin-dipole distributions measured via charge-exchange reactions. The calculations, which take into account the difference in deformation of the $\beta\beta$-decay mother and daughter nuclei, describe the measured GT and spin-dipole distributions reasonably well. The GT transition strengths from $^{150}$Nd to $^{150}$Pm at excitation energies below 2 MeV are too high in QRPA compared to the data, resulting in a $2\nu\beta\beta$ matrix element that exceeds the upper limit set by the data if full coherence between all contributions is assumed in both theory and data. However, the QRPA calculations are qualitatively consistent with the experimental finding that GT transitions to $1^{+}$ level in $^{150}$Pm contribute to the $2\nu\beta\beta$ matrix element.
The comparison between the acquired data and the theory can serve as a tool for future theoretical work, including establishing uncertainties in the estimates for $0\nu\beta\beta$ matrix elements. Such studies are important for large-scale direct counting experiments that aim to use $^{150}$Nd, such as SNO+\cite{KRA10}, SuperNEMO \cite{DEP10}, and DCBA \cite{ISH08}.

\begin{acknowledgments}
We thank the staff at NSCL and RCNP for their efforts in support of the $^{150}$Sm($t$,$^3$He) and $^{150}$Nd($^3$He,$t$) experiments. This work was supported by the US NSF (PHY-0822648 (JINA), PHY-0606007, and PHY-0758099). A.F.,D.F. and V.R. acknowledge the support of the Deutsche Forschungsgemeinschaft under both SFB TR27 ``Neutrinos and Beyond'' and Graduiertenkolleg GRK683.
\end{acknowledgments}

\bibliography{prc}

\begin{thebibliography}{93}
\expandafter\ifx\csname natexlab\endcsname\relax\def\natexlab#1{#1}\fi
\expandafter\ifx\csname bibnamefont\endcsname\relax
  \def\bibnamefont#1{#1}\fi
\expandafter\ifx\csname bibfnamefont\endcsname\relax
  \def\bibfnamefont#1{#1}\fi
\expandafter\ifx\csname citenamefont\endcsname\relax
  \def\citenamefont#1{#1}\fi
\expandafter\ifx\csname url\endcsname\relax
  \def\url#1{\texttt{#1}}\fi
\expandafter\ifx\csname urlprefix\endcsname\relax\def\urlprefix{URL }\fi
\providecommand{\bibinfo}[2]{#2}
\providecommand{\eprint}[2][]{\url{#2}}

\bibitem[{\citenamefont{Goeppert-Mayer}(1935)}]{GMAY35}
\bibinfo{author}{\bibfnamefont{M.}~\bibnamefont{Goeppert-Mayer}},
  \bibinfo{journal}{Phys. Rev.} \textbf{\bibinfo{volume}{\textbf{48}}},
  \bibinfo{pages}{512} (\bibinfo{year}{1935}).

\bibitem[{\citenamefont{Furry}(1939)}]{FUR39}
\bibinfo{author}{\bibfnamefont{W.}~\bibnamefont{Furry}},
  \bibinfo{journal}{Phys. Rev.} \textbf{\bibinfo{volume}{\textbf{56}}},
  \bibinfo{pages}{1184} (\bibinfo{year}{1939}).

\bibitem[{\citenamefont{{A}alseth{\emph{ et al.}}}(2004)}]{AAL04}
\bibinfo{author}{\bibfnamefont{C.}~\bibnamefont{{A}alseth{\emph{ et al.}}}},
  \bibinfo{journal}{ar{X}iv:hep-ph/0412300v1}  (\bibinfo{year}{2004}).

\bibitem[{\citenamefont{Mohapatra{\emph{ et al.}}}(2007)}]{MOH07}
\bibinfo{author}{\bibfnamefont{R.}~\bibnamefont{Mohapatra{\emph{ et al.}}}},
  \bibinfo{journal}{Rep. Prog. Phys.} \textbf{\bibinfo{volume}{\textbf{70}}},
  \bibinfo{pages}{1757} (\bibinfo{year}{2007}).

\bibitem[{\citenamefont{Avignone et~al.}(2008)\citenamefont{Avignone, Elliott,
  and Engel}}]{AVI08}
\bibinfo{author}{\bibfnamefont{F.~T.} \bibnamefont{Avignone}},
  \bibinfo{author}{\bibfnamefont{S.~R.} \bibnamefont{Elliott}},
  \bibnamefont{and} \bibinfo{author}{\bibfnamefont{J.}~\bibnamefont{Engel}},
  \bibinfo{journal}{Rev. Mod. Phys.} \textbf{\bibinfo{volume}{\textbf{80}}},
  \bibinfo{pages}{481} (\bibinfo{year}{2008}).

\bibitem[{\citenamefont{{T}omoda}(1991)}]{TOM91}
\bibinfo{author}{\bibfnamefont{T.}~\bibnamefont{{T}omoda}},
  \bibinfo{journal}{{R}ep. {P}rog. {P}hys.}
  \textbf{\bibinfo{volume}{\textbf{54}}}, \bibinfo{pages}{53}
  (\bibinfo{year}{1991}).

\bibitem[{\citenamefont{{M}uto}(1994)}]{MUT94}
\bibinfo{author}{\bibfnamefont{K.}~\bibnamefont{{M}uto}},
  \bibinfo{journal}{{N}ucl. {P}hys. {A}}
  \textbf{\bibinfo{volume}{\textbf{577}}}, \bibinfo{pages}{415c}
  (\bibinfo{year}{1994}).

\bibitem[{\citenamefont{{E}lliot and {V}ogel}(2002)}]{Ell02}
\bibinfo{author}{\bibfnamefont{S.}~\bibnamefont{{E}lliot}} \bibnamefont{and}
  \bibinfo{author}{\bibfnamefont{P.}~\bibnamefont{{V}ogel}},
  \bibinfo{journal}{{A}nn. {R}ev. {N}ucl. {P}art. {S}ci.}
  \textbf{\bibinfo{volume}{\textbf{52}}}, \bibinfo{pages}{115}
  (\bibinfo{year}{2002}).

\bibitem[{\citenamefont{Abad et~al.}(1984)\citenamefont{Abad, Morales,
  Nu{\~n}ez-Lagos, and Pacheco}}]{ABA84}
\bibinfo{author}{\bibfnamefont{J.}~\bibnamefont{Abad}},
  \bibinfo{author}{\bibfnamefont{A.}~\bibnamefont{Morales}},
  \bibinfo{author}{\bibfnamefont{R.}~\bibnamefont{Nu{\~n}ez-Lagos}},
  \bibnamefont{and} \bibinfo{author}{\bibfnamefont{A.~F.}
  \bibnamefont{Pacheco}}, \bibinfo{journal}{{{A}nales de {F}isica, {S}erie
  {A}}} \textbf{\bibinfo{volume}{\textbf{80}}}, \bibinfo{pages}{9}
  (\bibinfo{year}{1984}).

\bibitem[{\citenamefont{{{J}. {S}uhonen and {O}. {C}ivitarese}}(2000)}]{SUH00a}
\bibinfo{author}{\bibnamefont{{{J}. {S}uhonen and {O}. {C}ivitarese}}},
  \bibinfo{journal}{{C}zech. {J}ourn. {P}hys}
  \textbf{\bibinfo{volume}{\textbf{50}}}, \bibinfo{pages}{561}
  (\bibinfo{year}{2000}).

\bibitem[{\citenamefont{Ejiri and Toki}(1996)}]{EJI96}
\bibinfo{author}{\bibfnamefont{H.}~\bibnamefont{Ejiri}} \bibnamefont{and}
  \bibinfo{author}{\bibfnamefont{H.}~\bibnamefont{Toki}}, \bibinfo{journal}{J.
  Phys. Soc. Japan} \textbf{\bibinfo{volume}{\textbf{65}}}, \bibinfo{pages}{7}
  (\bibinfo{year}{1996}).

\bibitem[{\citenamefont{Ejiri}(2009)}]{EJI09}
\bibinfo{author}{\bibfnamefont{H.}~\bibnamefont{Ejiri}}, \bibinfo{journal}{J.
  Phys. Soc. Japan} \textbf{\bibinfo{volume}{\textbf{78}}},
  \bibinfo{pages}{074201} (\bibinfo{year}{2009}).

\bibitem[{\citenamefont{Ejiri}(2005)}]{EJI05}
\bibinfo{author}{\bibfnamefont{H.}~\bibnamefont{Ejiri}}, \bibinfo{journal}{J.
  Phys. Soc. Japan} \textbf{\bibinfo{volume}{\textbf{74}}},
  \bibinfo{pages}{2101} (\bibinfo{year}{2005}).

\bibitem[{\citenamefont{{{C}hristine {K}raus and {S}imon {J}.{M}.
  {P}eeters}}(2010)}]{KRA10}
\bibinfo{author}{\bibnamefont{{{C}hristine {K}raus and {S}imon {J}.{M}.
  {P}eeters}}}, \bibinfo{journal}{{P}rog. {N}ucl. {P}art. {P}hys.}
  \textbf{\bibinfo{volume}{\textbf{64}}}, \bibinfo{pages}{273}
  (\bibinfo{year}{2010}).

\bibitem[{\citenamefont{{{F}rank {D}eppisch, {C}hris {J}ackson, {I}rina Nasteva
  and Stefan S\"{o}ldner-Rembold}}(2010)}]{DEP10}
\bibinfo{author}{\bibnamefont{{{F}rank {D}eppisch, {C}hris {J}ackson, {I}rina
  Nasteva and Stefan S\"{o}ldner-Rembold}}}, \bibinfo{journal}{{P}rog. {N}ucl.
  {P}art. {P}hys.} \textbf{\bibinfo{volume}{\textbf{64}}}, \bibinfo{pages}{278}
  (\bibinfo{year}{2010}).

\bibitem[{\citenamefont{{N. Ishihara, T. Ohama, and Y. Yamada}}(1996)}]{ISH96}
\bibinfo{author}{\bibnamefont{{N. Ishihara, T. Ohama, and Y. Yamada}}},
  \bibinfo{journal}{Nucl. Instrum. Meth. Phys. Res. A}
  \textbf{\bibinfo{volume}{\textbf{373}}}, \bibinfo{pages}{325}
  (\bibinfo{year}{1996}).

\bibitem[{\citenamefont{Ishihara\emph{ et al.}}(2008)}]{ISH08}
\bibinfo{author}{\bibfnamefont{N.}~\bibnamefont{Ishihara\emph{ et al.}}},
  \bibinfo{journal}{J. Phys. Conf. Ser.}
  \textbf{\bibinfo{volume}{\textbf{120}}}, \bibinfo{pages}{052062}
  (\bibinfo{year}{2008}).

\bibitem[{\citenamefont{{D.L. {F}ang, Amand Faessler, Vadim Rodin,
  and Fedor \v{S}imkovic}}(2010)}]{FAN10a}
\bibinfo{author}{\bibnamefont{{D.L. {F}ang, Amand Faessler, Vadim
  Rodin, and Fedor \v{S}imkovic}}}, \bibinfo{journal}{{P}hys. {R}ev. {C}}
  \textbf{\bibinfo{volume}{\textbf{82}}}, \bibinfo{pages}{051301(R)}
  (\bibinfo{year}{2010}).

\bibitem[{\citenamefont{{D.L. {F}ang, Amand Faessler, Vadim Rodin,
  and Fedor \v{S}imkovic}}(2011)}]{FAN11}
\bibinfo{author}{\bibnamefont{{D.L. {F}ang, Amand Faessler, Vadim
  Rodin, and Fedor \v{S}imkovic}}}, \bibinfo{journal}{{P}hys. {R}ev. {C}}
  \textbf{\bibinfo{volume}{\textbf{83}}}, \bibinfo{pages}{034320}
  (\bibinfo{year}{2011}).

\bibitem[{\citenamefont{Barea and Iachello}(2009)}]{Bar09}
\bibinfo{author}{\bibfnamefont{J.}~\bibnamefont{Barea}} \bibnamefont{and}
  \bibinfo{author}{\bibfnamefont{F.}~\bibnamefont{Iachello}},
  \bibinfo{journal}{Phys. Rev. C} \textbf{\bibinfo{volume}{\textbf{79}}},
  \bibinfo{pages}{044301} (\bibinfo{year}{2009}).

\bibitem[{\citenamefont{Rodr\'{i}guez and Mart\'{i}nez-Pinedo}(2010)}]{Rodri10}
\bibinfo{author}{\bibfnamefont{T.~R.} \bibnamefont{Rodr\'{i}guez}}
  \bibnamefont{and}
  \bibinfo{author}{\bibfnamefont{G.}~\bibnamefont{Mart\'{i}nez-Pinedo}},
  \bibinfo{journal}{Phys. Rev. Lett.} \textbf{\bibinfo{volume}{\textbf{105}}},
  \bibinfo{pages}{252503} (\bibinfo{year}{2010}).

\bibitem[{\citenamefont{{{R}astislav {D}vornick\'y, {F}edor \v{S}imkovic, and
  {A}mand {F}aessler}}(2007)}]{DVO07}
\bibinfo{author}{\bibnamefont{{{R}astislav {D}vornick\'y, {F}edor \v{S}imkovic,
  and {A}mand {F}aessler}}}, in \emph{\bibinfo{booktitle}{{W}orkshop on
  {C}alculation of {D}ouble {B}eta {D}ecay {M}atrix {E}lements ({MEDEX}'07)}}
  (\bibinfo{publisher}{{AIP C}onf. {P}roc}, \bibinfo{year}{2007}), vol.
  \bibinfo{volume}{\textbf{942}}, pp. \bibinfo{pages}{28--32}.

\bibitem[{\citenamefont{Dermateosian and Tuli}(1995)}]{DER95}
\bibinfo{author}{\bibfnamefont{E.}~\bibnamefont{Dermateosian}}
  \bibnamefont{and} \bibinfo{author}{\bibfnamefont{J.~K.} \bibnamefont{Tuli}},
  \bibinfo{journal}{Nucl. Data Sheets} \textbf{\bibinfo{volume}{\textbf{75}}},
  \bibinfo{pages}{827} (\bibinfo{year}{1995}).

\bibitem[{\citenamefont{{K}olhinen{\emph{ et al.}}}(2010)}]{KOH10}
\bibinfo{author}{\bibfnamefont{V.}~\bibnamefont{{K}olhinen{\emph{ et al.}}}},
  \bibinfo{journal}{{P}hys. {R}ev. {C}} \textbf{\bibinfo{volume}{\textbf{82}}},
  \bibinfo{pages}{022501(R)} (\bibinfo{year}{2010}).

\bibitem[{\citenamefont{{Matthew Redshaw, Brianna J. Mount, Edmund G. Myers,
  and Frank T. Avignone, III}}(2009)}]{RED09}
\bibinfo{author}{\bibnamefont{{Matthew Redshaw, Brianna J. Mount, Edmund G.
  Myers, and Frank T. Avignone, III}}}, \bibinfo{journal}{{P}hys. {R}ev.
  {L}ett.} \textbf{\bibinfo{volume}{\textbf{102}}}, \bibinfo{pages}{212502}
  (\bibinfo{year}{2009}).

\bibitem[{\citenamefont{{Brianna J. Mount, Matthew Redshaw, and Edmund G.
  Myers}}(2010)}]{MOU10}
\bibinfo{author}{\bibnamefont{{Brianna J. Mount, Matthew Redshaw, and Edmund G.
  Myers}}}, \bibinfo{journal}{{P}hys. {R}ev. {C}}
  \textbf{\bibinfo{volume}{\textbf{81}}}, \bibinfo{pages}{032501(R)}
  (\bibinfo{year}{2010}).

\bibitem[{\citenamefont{{S. Rahaman, V.-V. Elomaa, T. Eronen, J. Hakala, A.
  Jokinen, A. Kankainen, J. Rissanen, J. Suhonen, C. Weber, and J.
  \"Ayst\"o}}(2009)}]{RAH09}
\bibinfo{author}{\bibnamefont{{S. Rahaman, V.-V. Elomaa, T. Eronen, J. Hakala,
  A. Jokinen, A. Kankainen, J. Rissanen, J. Suhonen, C. Weber, and J.
  \"Ayst\"o}}}, \bibinfo{journal}{{P}hys. {R}ev. {L}ett.}
  \textbf{\bibinfo{volume}{\textbf{103}}}, \bibinfo{pages}{042501}
  (\bibinfo{year}{2009}).

\bibitem[{\citenamefont{{S}chiffer{\emph{ et al.}}}(2008)}]{SCH08}
\bibinfo{author}{\bibfnamefont{J.}~\bibnamefont{{S}chiffer{\emph{ et al.}}}},
  \bibinfo{journal}{{P}hys. {R}ev. {L}ett.}
  \textbf{\bibinfo{volume}{\textbf{100}}}, \bibinfo{pages}{112501}
  (\bibinfo{year}{2008}).

\bibitem[{\citenamefont{{K}ay{\emph{ et al.}}}(2009)}]{KAY09}
\bibinfo{author}{\bibfnamefont{B.}~\bibnamefont{{K}ay{\emph{ et al.}}}},
  \bibinfo{journal}{{P}hys. {R}ev. {C}} \textbf{\bibinfo{volume}{\textbf{79}}},
  \bibinfo{pages}{021301} (\bibinfo{year}{2009}).

\bibitem[{\citenamefont{{D}ohmann{\emph{ et al.}}}(2008)}]{DOH08}
\bibinfo{author}{\bibfnamefont{H.}~\bibnamefont{{D}ohmann{\emph{ et al.}}}},
  \bibinfo{journal}{{P}hys. {R}ev. {C}} \textbf{\bibinfo{volume}{\textbf{78}}},
  \bibinfo{pages}{041602} (\bibinfo{year}{2008}).

\bibitem[{\citenamefont{{Y}ako{\emph{ et al.}}}(2009)}]{YAK09}
\bibinfo{author}{\bibfnamefont{K.}~\bibnamefont{{Y}ako{\emph{ et al.}}}},
  \bibinfo{journal}{{P}hys. {R}ev. {L}ett.}
  \textbf{\bibinfo{volume}{\textbf{103}}}, \bibinfo{pages}{012503}
  (\bibinfo{year}{2009}).

\bibitem[{\citenamefont{{G}rewe{\emph{ et al.}}}(2008{\natexlab{a}})}]{GRE08a}
\bibinfo{author}{\bibfnamefont{E.-W.} \bibnamefont{{G}rewe{\emph{ et al.}}}},
  \bibinfo{journal}{{P}hys. {R}ev. {C}} \textbf{\bibinfo{volume}{\textbf{78}}},
  \bibinfo{pages}{044301} (\bibinfo{year}{2008}{\natexlab{a}}).

\bibitem[{\citenamefont{{G}rewe{\emph{ et al.}}}(2008{\natexlab{b}})}]{GRE08}
\bibinfo{author}{\bibfnamefont{E.-W.} \bibnamefont{{G}rewe{\emph{ et al.}}}},
  \bibinfo{journal}{{P}hys. {R}ev. {C}} \textbf{\bibinfo{volume}{\textbf{77}}},
  \bibinfo{pages}{064303} (\bibinfo{year}{2008}{\natexlab{b}}).

\bibitem[{\citenamefont{{A}kimune et~al.}(1997)\citenamefont{{A}kimune, Ejiri,
  Fujiwara, Daito, Inomata, Hazama, Tamii, Toyokawa, and Yosoi}}]{AKI97}
\bibinfo{author}{\bibfnamefont{H.}~\bibnamefont{{A}kimune}},
  \bibinfo{author}{\bibfnamefont{H.}~\bibnamefont{Ejiri}},
  \bibinfo{author}{\bibfnamefont{M.}~\bibnamefont{Fujiwara}},
  \bibinfo{author}{\bibfnamefont{I.}~\bibnamefont{Daito}},
  \bibinfo{author}{\bibfnamefont{T.}~\bibnamefont{Inomata}},
  \bibinfo{author}{\bibfnamefont{R.}~\bibnamefont{Hazama}},
  \bibinfo{author}{\bibfnamefont{A.}~\bibnamefont{Tamii}},
  \bibinfo{author}{\bibfnamefont{H.}~\bibnamefont{Toyokawa}}, \bibnamefont{and}
  \bibinfo{author}{\bibfnamefont{M.}~\bibnamefont{Yosoi}},
  \bibinfo{journal}{{P}hys. {L}ett. {B}}
  \textbf{\bibinfo{volume}{\textbf{394}}}, \bibinfo{pages}{23}
  (\bibinfo{year}{1997}).

\bibitem[{\citenamefont{{R}akers{\emph{ et al.}}}(2004)}]{RAK04}
\bibinfo{author}{\bibfnamefont{S.}~\bibnamefont{{R}akers{\emph{ et al.}}}},
  \bibinfo{journal}{{P}hys. {R}ev. {C}} \textbf{\bibinfo{volume}{\textbf{70}}},
  \bibinfo{pages}{054302} (\bibinfo{year}{2004}).

\bibitem[{\citenamefont{{G}rewe{\emph{ et al.}}}(2007)}]{GRE07}
\bibinfo{author}{\bibfnamefont{E.-W.} \bibnamefont{{G}rewe{\emph{ et al.}}}},
  \bibinfo{journal}{{P}hys. {R}ev. {C}} \textbf{\bibinfo{volume}{\textbf{76}}},
  \bibinfo{pages}{054307} (\bibinfo{year}{2007}).

\bibitem[{\citenamefont{Ejiri}(2000)}]{EJI00}
\bibinfo{author}{\bibfnamefont{H.}~\bibnamefont{Ejiri}},
  \bibinfo{journal}{Phys. Rep.} \textbf{\bibinfo{volume}{\textbf{338}}},
  \bibinfo{pages}{265} (\bibinfo{year}{2000}).

\bibitem[{\citenamefont{{M. S. Yousef, V. Rodin, A. Faessler, and F.
  \v{S}imkovic}}(2009)}]{YOU09}
\bibinfo{author}{\bibnamefont{{M. S. Yousef, V. Rodin, A. Faessler, and F.
  \v{S}imkovic}}}, \bibinfo{journal}{Phys. Rev. C}
  \textbf{\bibinfo{volume}{\textbf{79}}}, \bibinfo{pages}{014314}
  (\bibinfo{year}{2009}).

\bibitem[{\citenamefont{{D.L. {F}ang, Amand Faessler, Vadim Rodin,
  Mohamed Saleh Yousef, and Fedor \v{S}imkovic}}(2010)}]{FAN10}
\bibinfo{author}{\bibnamefont{{D.L. {F}ang, Amand Faessler, Vadim
  Rodin, Mohamed Saleh Yousef, and Fedor \v{S}imkovic}}},
  \bibinfo{journal}{{P}hys. {R}ev. {C}} \textbf{\bibinfo{volume}{\textbf{81}}},
  \bibinfo{pages}{037303} (\bibinfo{year}{2010}).

\bibitem[{\citenamefont{Harakeh and van~der Woude}(2001)}]{HAR01}
\bibinfo{author}{\bibfnamefont{M.~N.} \bibnamefont{Harakeh}} \bibnamefont{and}
  \bibinfo{author}{\bibfnamefont{A.}~\bibnamefont{van~der Woude}},
  \emph{\bibinfo{title}{Giant Resonances: Fundamental High-Frequency Modes of
  Nuclear Excitations}} (\bibinfo{publisher}{Oxford University Press},
  \bibinfo{address}{New York}, \bibinfo{year}{2001}).

\bibitem[{\citenamefont{{L}ove and {F}raney}(1981)}]{LOV81}
\bibinfo{author}{\bibfnamefont{W.~G.}~\bibnamefont{{L}ove}} \bibnamefont{and}
  \bibinfo{author}{\bibfnamefont{M.~A.}~\bibnamefont{{F}raney}},
  \bibinfo{journal}{{P}hys. {R}ev. {C}} \textbf{\bibinfo{volume}{\textbf{24}}},
  \bibinfo{pages}{1073} (\bibinfo{year}{1981}).

\bibitem[{\citenamefont{Franey and Love}(1985)}]{FRA85}
\bibinfo{author}{\bibfnamefont{M.~A.} \bibnamefont{Franey}} \bibnamefont{and}
  \bibinfo{author}{\bibfnamefont{W.~G.} \bibnamefont{Love}},
  \bibinfo{journal}{Phys. Rev. C} \textbf{\bibinfo{volume}{\textbf{31}}},
  \bibinfo{pages}{488} (\bibinfo{year}{1985}).

\bibitem[{\citenamefont{{J. Rapaport and E. Sugarbaker}}(1994)}]{RAP94}
\bibinfo{author}{\bibnamefont{{J. Rapaport and E. Sugarbaker}}},
  \bibinfo{journal}{Annu. Rev. Nucl. Part. Sci.}
  \textbf{\bibinfo{volume}{\textbf{44}}}, \bibinfo{pages}{109}
  (\bibinfo{year}{1994}).

\bibitem[{\citenamefont{{T}addeucci et~al.}(1987)\citenamefont{{T}addeucci,
  {G}oulding, {C}arey, {B}yrd, {G}oodman, {G}aarde, {L}arsen, {H}oren,
  {R}apaport, and {S}ugarbaker}}]{TAD87}
\bibinfo{author}{\bibfnamefont{T.~N.} \bibnamefont{{T}addeucci}},
  \bibinfo{author}{\bibfnamefont{C.}~\bibnamefont{{G}oulding}},
  \bibinfo{author}{\bibfnamefont{T.}~\bibnamefont{{C}arey}},
  \bibinfo{author}{\bibfnamefont{R.}~\bibnamefont{{B}yrd}},
  \bibinfo{author}{\bibfnamefont{C.}~\bibnamefont{{G}oodman}},
  \bibinfo{author}{\bibfnamefont{C.}~\bibnamefont{{G}aarde}},
  \bibinfo{author}{\bibfnamefont{J.}~\bibnamefont{{L}arsen}},
  \bibinfo{author}{\bibfnamefont{D.}~\bibnamefont{{H}oren}},
  \bibinfo{author}{\bibfnamefont{J.}~\bibnamefont{{R}apaport}},
  \bibnamefont{and}
  \bibinfo{author}{\bibfnamefont{E.}~\bibnamefont{{S}ugarbaker}},
  \bibinfo{journal}{{N}ucl. {P}hys. {A}}
  \textbf{\bibinfo{volume}{\textbf{469}}}, \bibinfo{pages}{125}
  (\bibinfo{year}{1987}).

\bibitem[{\citenamefont{Zegers{\emph{ et al.}}}(2007)}]{ZEG07}
\bibinfo{author}{\bibfnamefont{R.~G.~T.} \bibnamefont{Zegers{\emph{ et al.}}}},
  \bibinfo{journal}{{P}hys. {R}ev. {L}ett.}
  \textbf{\bibinfo{volume}{\textbf{99}}}, \bibinfo{pages}{202501}
  (\bibinfo{year}{2007}).

\bibitem[{\citenamefont{Perdikakis{\emph{ et al.}}}()}]{PER11}
\bibinfo{author}{\bibfnamefont{G.}~\bibnamefont{Perdikakis{\emph{ et al.}}}},
  \bibinfo{note}{to be published.}

\bibitem[{\citenamefont{Ellegaard{\emph{ et al.}}}(1983)}]{ELL83}
\bibinfo{author}{\bibfnamefont{C.}~\bibnamefont{Ellegaard{\emph{ et al.}}}},
  \bibinfo{journal}{Phys. Rev. Lett.} \textbf{\bibinfo{volume}{\textbf{50}}},
  \bibinfo{pages}{1745} (\bibinfo{year}{1983}).

\bibitem[{\citenamefont{{N. Auerbach and A. Klein}}(1983)}]{AUE83}
\bibinfo{author}{\bibnamefont{{N. Auerbach and A. Klein}}},
  \bibinfo{journal}{{N}ucl. {P}hys. {A}}
  \textbf{\bibinfo{volume}{\textbf{395}}}, \bibinfo{pages}{77}
  (\bibinfo{year}{1983}).

\bibitem[{\citenamefont{Zegers{\emph{ et al.}}}(2000)}]{ZEG00}
\bibinfo{author}{\bibfnamefont{R.~G.~T.} \bibnamefont{Zegers{\emph{ et al.}}}},
  \bibinfo{journal}{Phys. Rev. Lett.} \textbf{\bibinfo{volume}{\textbf{84}}},
  \bibinfo{pages}{3779} (\bibinfo{year}{2000}).

\bibitem[{\citenamefont{Zegers et~al.}(2001)\citenamefont{Zegers, van~den Berg,
  Brandenburg, Fujiwara, Guillot, Harakeh, Laurent, van~der Werf, Willis, and
  Wilschut}}]{ZEG01}
\bibinfo{author}{\bibfnamefont{R.~G.~T.} \bibnamefont{Zegers}},
  \bibinfo{author}{\bibfnamefont{A.~M.} \bibnamefont{van~den Berg}},
  \bibinfo{author}{\bibfnamefont{S.}~\bibnamefont{Brandenburg}},
  \bibinfo{author}{\bibfnamefont{M.}~\bibnamefont{Fujiwara}},
  \bibinfo{author}{\bibfnamefont{J.}~\bibnamefont{Guillot}},
  \bibinfo{author}{\bibfnamefont{M.~N.}~\bibnamefont{Harakeh}},
  \bibinfo{author}{\bibfnamefont{H.}~\bibnamefont{Laurent}},
  \bibinfo{author}{\bibfnamefont{S.~Y.}~\bibnamefont{van~der Werf}},
  \bibinfo{author}{\bibfnamefont{A.}~\bibnamefont{Willis}}, \bibnamefont{and}
  \bibinfo{author}{\bibfnamefont{H.~W.}~\bibnamefont{Wilschut}},
  \bibinfo{journal}{Phys. Rev. C} \textbf{\bibinfo{volume}{\textbf{63}}},
  \bibinfo{pages}{034613} (\bibinfo{year}{2001}).

\bibitem[{\citenamefont{Zegers{\emph{ et al.}}}(2003)}]{ZEG03}
\bibinfo{author}{\bibfnamefont{R.~G.~T.} \bibnamefont{Zegers{\emph{ et al.}}}},
  \bibinfo{journal}{Phys. Rev. Lett.} \textbf{\bibinfo{volume}{\textbf{90}}},
  \bibinfo{pages}{202501} (\bibinfo{year}{2003}).

\bibitem[{\citenamefont{Auerbach and Klein}(1984)}]{AUE84}
\bibinfo{author}{\bibfnamefont{N.}~\bibnamefont{Auerbach}} \bibnamefont{and}
  \bibinfo{author}{\bibfnamefont{A.}~\bibnamefont{Klein}},
  \bibinfo{journal}{Phys. Rev. C} \textbf{\bibinfo{volume}{\textbf{30}}},
  \bibinfo{pages}{1032} (\bibinfo{year}{1984}).

\bibitem[{\citenamefont{Guillot{\emph{ et al.}}}(2006)}]{GUI06}
\bibinfo{author}{\bibfnamefont{J.}~\bibnamefont{Guillot{\emph{ et al.}}}},
  \bibinfo{journal}{Phys. Rev. C} \textbf{\bibinfo{volume}{\textbf{73}}},
  \bibinfo{pages}{014616} (\bibinfo{year}{2006}).

\bibitem[{\citenamefont{Yako{\emph{ et al.}}}(2005)}]{YAK05}
\bibinfo{author}{\bibfnamefont{K.}~\bibnamefont{Yako{\emph{ et al.}}}},
  \bibinfo{journal}{Phys. Lett.} \textbf{\bibinfo{volume}{\textbf{B615}}},
  \bibinfo{pages}{193} (\bibinfo{year}{2005}).

\bibitem[{\citenamefont{Auerbach et~al.}(1989)\citenamefont{Auerbach,
  Osterfeld, and Udagawa}}]{AUE89}
\bibinfo{author}{\bibfnamefont{N.}~\bibnamefont{Auerbach}},
  \bibinfo{author}{\bibfnamefont{F.}~\bibnamefont{Osterfeld}},
  \bibnamefont{and} \bibinfo{author}{\bibfnamefont{T.}~\bibnamefont{Udagawa}},
  \bibinfo{journal}{Phys. Lett.} \textbf{\bibinfo{volume}{\bf{B219}}},
  \bibinfo{pages}{184} (\bibinfo{year}{1989}).

\bibitem[{\citenamefont{Auerbach}(1998)}]{AUE98}
\bibinfo{author}{\bibfnamefont{N.}~\bibnamefont{Auerbach}},
  \bibinfo{journal}{Comm. Nucl. Part. Phys.}
  \textbf{\bibinfo{volume}{\bf{22}}}, \bibinfo{pages}{223}
  (\bibinfo{year}{1998}).

\bibitem[{\citenamefont{{F}ujiwara{\emph{ et al.}}}(1999)}]{FUJ99}
\bibinfo{author}{\bibfnamefont{M.}~\bibnamefont{{F}ujiwara{\emph{ et al.}}}},
  \bibinfo{journal}{{N}ucl. {I}nstrum. {M}eth. {P}hys. {R}es. {A}}
  \textbf{\bibinfo{volume}{\textbf{422}}}, \bibinfo{pages}{484}
  (\bibinfo{year}{1999}).

\bibitem[{\citenamefont{{F}ujita{\emph{ et al.}}}(2002)}]{FUJ02}
\bibinfo{author}{\bibfnamefont{H.}~\bibnamefont{{F}ujita{\emph{ et al.}}}},
  \bibinfo{journal}{{N}ucl. {I}nstrum. {M}eth. {P}hys. {R}es. {A}}
  \textbf{\bibinfo{volume}{\textbf{484}}}, \bibinfo{pages}{17}
  (\bibinfo{year}{2002}).

\bibitem[{\citenamefont{{F}ujita{\emph{ et al.}}}(2001)}]{FUJ01}
\bibinfo{author}{\bibfnamefont{H.}~\bibnamefont{{F}ujita{\emph{ et al.}}}},
  \bibinfo{journal}{{N}ucl. {I}nstrum. {M}eth. {P}hys. {R}es. {A}}
  \textbf{\bibinfo{volume}{\textbf{469}}}, \bibinfo{pages}{55}
  (\bibinfo{year}{2001}).

\bibitem[{\citenamefont{Zegers{\emph{ et al.}}}(2004)}]{ZEG04}
\bibinfo{author}{\bibfnamefont{R.~G.~T.} \bibnamefont{Zegers{\emph{ et al.}}}},
  \bibinfo{journal}{Nucl. Phys. A} \textbf{\bibinfo{volume}{\textbf{731}}},
  \bibinfo{pages}{121c} (\bibinfo{year}{2004}).

\bibitem[{\citenamefont{Zegers{\emph{ et al.}}}(2006)}]{ZEG06}
\bibinfo{author}{\bibfnamefont{R.~G.~T.} \bibnamefont{Zegers{\emph{ et al.}}}},
  \bibinfo{journal}{{P}hys. {R}ev. {C}} \textbf{\bibinfo{volume}{\textbf{74}}},
  \bibinfo{pages}{024309} (\bibinfo{year}{2006}).

\bibitem[{\citenamefont{{R}aynal}()}]{ECIS}
\bibinfo{author}{\bibfnamefont{J.}~\bibnamefont{{R}aynal}},
  \bibinfo{note}{{ECIS-97} (unpublished)}.

\bibitem[{\citenamefont{{{C}.{J}. {G}uess{\emph{ et al.}}}}(2009)}]{GUE09}
\bibinfo{author}{\bibnamefont{{{C}.{J}. {G}uess{\emph{ et al.}}}}},
  \bibinfo{journal}{{P}hys. {R}ev. {C}} \textbf{\bibinfo{volume}{\textbf{80}}},
  \bibinfo{pages}{024305} (\bibinfo{year}{2009}).

\bibitem[{CCF()}]{CCF88}
\bibinfo{note}{The {K}500$\otimes${K}1200, a coupled cyclotron facility at the
  {NSCL}, {NSCL R}eport {MSUCL}-939, 1998}.

\bibitem[{\citenamefont{Morrissey et~al.}(2003)\citenamefont{Morrissey,
  Sherrill, Steiner, Stolz, , and Wiedenhoever}}]{MOR03}
\bibinfo{author}{\bibfnamefont{D.}~\bibnamefont{Morrissey}},
  \bibinfo{author}{\bibfnamefont{B.}~\bibnamefont{Sherrill}},
  \bibinfo{author}{\bibfnamefont{M.}~\bibnamefont{Steiner}},
  \bibinfo{author}{\bibfnamefont{A.}~\bibnamefont{Stolz}}, , \bibnamefont{and}
  \bibinfo{author}{\bibfnamefont{I.}~\bibnamefont{Wiedenhoever}},
  \bibinfo{journal}{{N}ucl. {I}nstrum. {M}eth. {P}hys. {R}es. {B}}
  \textbf{\bibinfo{volume}{\textbf{204}}}, \bibinfo{pages}{90}
  (\bibinfo{year}{2003}).

\bibitem[{\citenamefont{Hitt{\emph{ et al.}}}(2006)}]{HIT06}
\bibinfo{author}{\bibfnamefont{G.~W.} \bibnamefont{Hitt{\emph{ et al.}}}},
  \bibinfo{journal}{Nucl. Instrum. Meth. Phys. Res. A}
  \textbf{\bibinfo{volume}{\textbf{566}}}, \bibinfo{pages}{264}
  (\bibinfo{year}{2006}).

\bibitem[{\citenamefont{Bazin et~al.}(2003)\citenamefont{Bazin, Caggiano,
  Sherrill, Yurkon, and Zeller}}]{BAZ03}
\bibinfo{author}{\bibfnamefont{D.}~\bibnamefont{Bazin}},
  \bibinfo{author}{\bibfnamefont{J.~A.} \bibnamefont{Caggiano}},
  \bibinfo{author}{\bibfnamefont{B.~M.} \bibnamefont{Sherrill}},
  \bibinfo{author}{\bibfnamefont{J.}~\bibnamefont{Yurkon}}, \bibnamefont{and}
  \bibinfo{author}{\bibfnamefont{A.}~\bibnamefont{Zeller}},
  \bibinfo{journal}{Nucl. Instr. Meth. Phys. Res. B}
  \textbf{\bibinfo{volume}{\textbf{204}}}, \bibinfo{pages}{629}
  (\bibinfo{year}{2003}).

\bibitem[{\citenamefont{Yurkon et~al.}(1999)\citenamefont{Yurkon, Bazin,
  Benenson, Morrissey, Sherrill, Swan, and Swanson}}]{YUR99}
\bibinfo{author}{\bibfnamefont{J.}~\bibnamefont{Yurkon}},
  \bibinfo{author}{\bibfnamefont{D.}~\bibnamefont{Bazin}},
  \bibinfo{author}{\bibfnamefont{W.}~\bibnamefont{Benenson}},
  \bibinfo{author}{\bibfnamefont{D.~J.} \bibnamefont{Morrissey}},
  \bibinfo{author}{\bibfnamefont{B.~M.} \bibnamefont{Sherrill}},
  \bibinfo{author}{\bibfnamefont{D.}~\bibnamefont{Swan}}, \bibnamefont{and}
  \bibinfo{author}{\bibfnamefont{R.}~\bibnamefont{Swanson}},
  \bibinfo{journal}{Nucl. Instr. Meth. Phys. Res. A}
  \textbf{\bibinfo{volume}{\textbf{422}}}, \bibinfo{pages}{291}
  (\bibinfo{year}{1999}).

\bibitem[{\citenamefont{Makino and Berz}(1999)}]{COSY}
\bibinfo{author}{\bibfnamefont{K.}~\bibnamefont{Makino}} \bibnamefont{and}
  \bibinfo{author}{\bibfnamefont{M.}~\bibnamefont{Berz}},
  \bibinfo{journal}{Nucl. Instrum. Meth. Phys. Res. A}
  \textbf{\bibinfo{volume}{\textbf{427}}}, \bibinfo{pages}{338}
  (\bibinfo{year}{1999}).

\bibitem[{\citenamefont{{G. W. Hitt{\emph{ et al.}}}}(2009)}]{HIT09}
\bibinfo{author}{\bibnamefont{{G. W. Hitt{\emph{ et al.}}}}},
  \bibinfo{journal}{{P}hys. {R}ev. {C}}
  \textbf{\bibinfo{volume}{{\textbf{80}}}}, \bibinfo{pages}{014313}
  (\bibinfo{year}{2009}).

\bibitem[{\citenamefont{{M}oinester}(1987)}]{MON87}
\bibinfo{author}{\bibfnamefont{M.}~\bibnamefont{{M}oinester}},
  \bibinfo{journal}{{C}an. {J}ourn. of {P}hys.}
  \textbf{\bibinfo{volume}{\textbf{65}}}, \bibinfo{pages}{660}
  (\bibinfo{year}{1987}).

\bibitem[{\citenamefont{Cook and Carr}()}]{FOLD}
\bibinfo{author}{\bibfnamefont{J.}~\bibnamefont{Cook}} \bibnamefont{and}
  \bibinfo{author}{\bibfnamefont{J.}~\bibnamefont{Carr}},
  \bibinfo{note}{computer program \textsc{fold}, Florida State University
  (unpublished), based on F. Petrovich and D. Stanley, Nucl. Phys.
  {\textbf{A275}}, 487 (1977), modified as described in J. Cook {\emph{ et
  al.}}, Phys. Rev. C {\textbf{30}}, 1538 (1984) and R. G. T. Zegers, S.
  Fracasso and G. Col\`{o} (2006), unpublished.}

\bibitem[{\citenamefont{Hofstee{\emph{ et al.}}}(1995)}]{HOF95}
\bibinfo{author}{\bibfnamefont{M.~A.} \bibnamefont{Hofstee{\emph{ et al.}}}},
  \bibinfo{journal}{Nucl. Phys. A} \textbf{\bibinfo{volume}{\textbf{588}}},
  \bibinfo{pages}{729} (\bibinfo{year}{1995}).

\bibitem[{\citenamefont{{S. Y. van der Werf}}()}]{NORM}
\bibinfo{author}{\bibnamefont{{S. Y. van der Werf}}}, \bibinfo{note}{computer
  program \textsc{NORMOD}, unpublished}.

\bibitem[{\citenamefont{{B}rown}(1998)}]{BRO98}
\bibinfo{author}{\bibfnamefont{B.~A.} \bibnamefont{{B}rown}},
  \bibinfo{journal}{{P}hys. {R}ev. {C}} \textbf{\bibinfo{volume}{\textbf{58}}},
  \bibinfo{pages}{220} (\bibinfo{year}{1998}).

\bibitem[{\citenamefont{Pieper and Wiringa}(2001)}]{WIR05}
\bibinfo{author}{\bibfnamefont{S.~C.} \bibnamefont{Pieper}} \bibnamefont{and}
  \bibinfo{author}{\bibfnamefont{R.~B.} \bibnamefont{Wiringa}},
  \bibinfo{journal}{Annu. Rev. Nucl. Part. Sci.}
  \textbf{\bibinfo{volume}{\textbf{51}}}, \bibinfo{pages}{53}
  (\bibinfo{year}{2001}), \bibinfo{note}{and {R}.{B}. {W}iringa, private
  communication}.

\bibitem[{\citenamefont{{{S}.{Y}. van der {W}erf, {S}. {B}randenburg, {P}.
  {G}rasdijk, {W}.{A}. {S}terrenburg, {M}.{N}. {H}arakeh, {M}.{B}.
  {G}reenfield, {B}.{A}. {B}rown and {M}. {F}ujiwara}}(1989)}]{WER89}
\bibinfo{author}{\bibnamefont{{{S}.{Y}. van der {W}erf, {S}. {B}randenburg,
  {P}. {G}rasdijk, {W}.{A}. {S}terrenburg, {M}.{N}. {H}arakeh, {M}.{B}.
  {G}reenfield, {B}.{A}. {B}rown and {M}. {F}ujiwara}}},
  \bibinfo{journal}{{N}ucl. {P}hys. {A}}
  \textbf{\bibinfo{volume}{\textbf{496}}}, \bibinfo{pages}{305}
  (\bibinfo{year}{1989}).

\bibitem[{\citenamefont{Barrette et~al.}(1970)\citenamefont{Barrette, Barrette,
  Monaro, Santhanam, and Markiza}}]{BAR69}
\bibinfo{author}{\bibfnamefont{J.}~\bibnamefont{Barrette}},
  \bibinfo{author}{\bibfnamefont{M.}~\bibnamefont{Barrette}},
  \bibinfo{author}{\bibfnamefont{S.}~\bibnamefont{Monaro}},
  \bibinfo{author}{\bibfnamefont{S.}~\bibnamefont{Santhanam}},
  \bibnamefont{and} \bibinfo{author}{\bibfnamefont{S.}~\bibnamefont{Markiza}},
  \bibinfo{journal}{Can. J. Phys.} \textbf{\bibinfo{volume}{\textbf{48}}},
  \bibinfo{pages}{1161} (\bibinfo{year}{1970}).

\bibitem[{\citenamefont{J{\"{a}}necke{\emph{ et al.}}}(1993)}]{JAN93}
\bibinfo{author}{\bibfnamefont{J.}~\bibnamefont{J{\"{a}}necke{\emph{ et
  al.}}}}, \bibinfo{journal}{Phys. Rev. C}
  \textbf{\bibinfo{volume}{\textbf{48}}}, \bibinfo{pages}{2828}
  (\bibinfo{year}{1993}).

\bibitem[{\citenamefont{Pham{\emph{ et al.}}}(1995)}]{PHA95}
\bibinfo{author}{\bibfnamefont{K.}~\bibnamefont{Pham{\emph{ et al.}}}},
  \bibinfo{journal}{Phys. Rev. C} \textbf{\bibinfo{volume}{\textbf{51}}},
  \bibinfo{pages}{526} (\bibinfo{year}{1995}).

\bibitem[{\citenamefont{{V. G. Guba, M. A. Nikolaev and M. G.
  Urin}}(1989)}]{GUB89}
\bibinfo{author}{\bibnamefont{{V. G. Guba, M. A. Nikolaev and M. G. Urin}}},
  \bibinfo{journal}{{P}hys. {L}ett. {B}}
  \textbf{\bibinfo{volume}{\textbf{218}}}, \bibinfo{pages}{283}
  (\bibinfo{year}{1989}).

\bibitem[{\citenamefont{{{V}. {R}odin and {M}. {U}rin}}(2003)}]{ROD03}
\bibinfo{author}{\bibnamefont{{{V}. {R}odin and {M}. {U}rin}}},
  \bibinfo{journal}{{P}hys. {A}tom. {N}ucl.}
  \textbf{\bibinfo{volume}{\textbf{66}}}, \bibinfo{pages}{2128}
  (\bibinfo{year}{2003}).

\bibitem[{\citenamefont{Sasano{\emph{ et al.}}}(2009)}]{SAS09}
\bibinfo{author}{\bibfnamefont{M.}~\bibnamefont{Sasano{\emph{ et al.}}}},
  \bibinfo{journal}{AIP Conf. Proc.} \textbf{\bibinfo{volume}{\textbf{1180}}},
  \bibinfo{pages}{102} (\bibinfo{year}{2009}).

\bibitem[{\citenamefont{Gaarde et~al.}(1981)\citenamefont{Gaarde, Rapaport,
  Taddeucci, Goodman, Foster, Bainum, Goulding, Greenfield, H{\"{o}}ren, and
  Sugarbaker}}]{GAA81}
\bibinfo{author}{\bibfnamefont{C.}~\bibnamefont{Gaarde}},
  \bibinfo{author}{\bibfnamefont{J.}~\bibnamefont{Rapaport}},
  \bibinfo{author}{\bibfnamefont{T.~N.} \bibnamefont{Taddeucci}},
  \bibinfo{author}{\bibfnamefont{C.~D.} \bibnamefont{Goodman}},
  \bibinfo{author}{\bibfnamefont{C.~C.} \bibnamefont{Foster}},
  \bibinfo{author}{\bibfnamefont{D.~E.} \bibnamefont{Bainum}},
  \bibinfo{author}{\bibfnamefont{C.~A.} \bibnamefont{Goulding}},
  \bibinfo{author}{\bibfnamefont{M.~B.} \bibnamefont{Greenfield}},
  \bibinfo{author}{\bibfnamefont{D.~J.} \bibnamefont{H{\"{o}}ren}},
  \bibnamefont{and}
  \bibinfo{author}{\bibfnamefont{E.}~\bibnamefont{Sugarbaker}},
  \bibinfo{journal}{Nucl. Phys. A} \textbf{\bibinfo{volume}{\textbf{369}}},
  \bibinfo{pages}{258} (\bibinfo{year}{1981}).

\bibitem[{\citenamefont{Gaarde}(1985)}]{GAA85}
\bibinfo{author}{\bibfnamefont{C.}~\bibnamefont{Gaarde}}, in
  \emph{\bibinfo{booktitle}{Proc. Niels Bohr Centennial Conference on Nuclear
  Structure, Copenhagen}}, edited by \bibinfo{editor}{\bibfnamefont{R.~A.}
  \bibnamefont{Broglia}}, \bibinfo{editor}{\bibfnamefont{G.~B.}
  \bibnamefont{Hagemann}}, \bibnamefont{and}
  \bibinfo{editor}{\bibfnamefont{B.}~\bibnamefont{Herskind}}
  (\bibinfo{publisher}{North-Holland, Amsterdam}, \bibinfo{year}{1985}), p.
  \bibinfo{pages}{449c}.

\bibitem[{\citenamefont{Hyuga et~al.}(1980)\citenamefont{Hyuga, Arima, and
  Shimizu}}]{HYU80}
\bibinfo{author}{\bibfnamefont{H.}~\bibnamefont{Hyuga}},
  \bibinfo{author}{\bibfnamefont{A.}~\bibnamefont{Arima}}, \bibnamefont{and}
  \bibinfo{author}{\bibfnamefont{K.}~\bibnamefont{Shimizu}},
  \bibinfo{journal}{{N}ucl. {P}hys. {A}}
  \textbf{\bibinfo{volume}{\textbf{336}}}, \bibinfo{pages}{363}
  (\bibinfo{year}{1980}).

\bibitem[{\citenamefont{{A}rima}(1999)}]{ARI99}
\bibinfo{author}{\bibfnamefont{A.}~\bibnamefont{{A}rima}},
  \bibinfo{journal}{{N}ucl. {P}hys. {A}}
  \textbf{\bibinfo{volume}{\textbf{649}}}, \bibinfo{pages}{260c}
  (\bibinfo{year}{1999}).

\bibitem[{\citenamefont{{H}elmer{\emph{ et al.}}}(1997)}]{HEL97}
\bibinfo{author}{\bibfnamefont{R.}~\bibnamefont{{H}elmer{\emph{ et al.}}}},
  \bibinfo{journal}{{P}hys. {R}ev. {C}} \textbf{\bibinfo{volume}{\textbf{55}}},
  \bibinfo{pages}{2802} (\bibinfo{year}{1997}).

\bibitem[{\citenamefont{{R}aywood et~al.}(1997)\citenamefont{{R}aywood, {L}ong,
  and {S}picer}}]{RAY97}
\bibinfo{author}{\bibfnamefont{K.}~\bibnamefont{{R}aywood}},
  \bibinfo{author}{\bibfnamefont{S.}~\bibnamefont{{L}ong}}, \bibnamefont{and}
  \bibinfo{author}{\bibfnamefont{B.}~\bibnamefont{{S}picer}},
  \bibinfo{journal}{{N}ucl. {P}hys. {A}}
  \textbf{\bibinfo{volume}{\textbf{625}}}, \bibinfo{pages}{675}
  (\bibinfo{year}{1997}).

\bibitem[{\citenamefont{{R}akers{\emph{ et al.}}}(2005)}]{RAK05}
\bibinfo{author}{\bibfnamefont{S.}~\bibnamefont{{R}akers{\emph{ et al.}}}},
  \bibinfo{journal}{{P}hys. {R}ev. {C}} \textbf{\bibinfo{volume}{\textbf{71}}},
  \bibinfo{pages}{054313} (\bibinfo{year}{2005}).

\bibitem[{\citenamefont{{{J}. {S}uhonen and {O}. Civitarese}}(1998)}]{SUH98}
\bibinfo{author}{\bibnamefont{{{J}. {S}uhonen and {O}. Civitarese}}},
  \bibinfo{journal}{{P}hys. {R}ep.} \textbf{\bibinfo{volume}{\textbf{300}}},
  \bibinfo{pages}{123} (\bibinfo{year}{1998}).

\bibitem[{\citenamefont{{B}arabash}(2010)}]{BAR10}
\bibinfo{author}{\bibfnamefont{A.~S.}~\bibnamefont{{B}arabash}},
  \bibinfo{journal}{{P}hys. {R}ev. {C}} \textbf{\bibinfo{volume}{\textbf{81}}},
  \bibinfo{pages}{035501} (\bibinfo{year}{2010}).

\bibitem[{\citenamefont{{P}ritychenko}(2010)}]{PRI10}
\bibinfo{author}{\bibfnamefont{B.}~\bibnamefont{{P}ritychenko}},
  \bibinfo{journal}{ar{X}iv:1004.3280v1 [nucl-th]}  (\bibinfo{year}{2010}),
  \bibinfo{note}{{B}rookhaven {N}ational {L}aboratory {R}eport
  {BNL}-91299-2010}.

\end{thebibliography}


\end{document}